\documentclass[a4paper,fleqn,usenatbib]{mnras}

\usepackage{newtxtext,newtxmath}

\usepackage[T1]{fontenc}
\usepackage{ae,aecompl}

\usepackage{graphicx}	
\usepackage{amsmath}	
\usepackage{amssymb}	
\usepackage{tabularx}
\usepackage{float}
\restylefloat{table}
\usepackage[labelfont=bf]{caption}
\usepackage{natbib}
\usepackage[caption=false]{subfig}
\usepackage{placeins}

\title[Dependence of Halo Mass on Galaxy Size]{The Dependence of Halo Mass on Galaxy Size at Fixed Stellar Mass Using Weak Lensing}

\author[Paul Charlton et al.]{
Paul J. L. Charlton,$^{1}$\thanks{E-mail:pcharlton@uwaterloo.ca}
Michael J. Hudson,$^{1,2}$
Michael L. Balogh,$^{1}$
and Sumeet Khatri$^{1}$
\\
$^{1}$Department of Physics and Astronomy, University of Waterloo, Waterloo, ON N2L 3G1, Canada.\\
$^{2}$Perimeter Institute for Theoretical Physics, 31 Caroline St. N., Waterloo, ON, N2L 2Y5, Canada.\\
}

\date{Accepted XXX. Received YYY; in original form ZZZ}

\pubyear{2017}

\begin{document}
\label{firstpage}
\pagerange{\pageref{firstpage}--\pageref{lastpage}}
\maketitle

\begin{abstract}
Stellar mass has been shown to correlate with halo mass, with non-negligible scatter. The stellar mass-size and luminosity-size relationships of galaxies also show significant scatter in galaxy size at fixed stellar mass. It is possible that, at fixed stellar mass and galaxy colour, the halo mass is correlated with galaxy size. Galaxy-galaxy lensing allows us to measure the mean masses of dark matter haloes for stacked samples of galaxies. We extend the analysis of the galaxies in the CFHTLenS catalogue by fitting single S\'{e}rsic surface brightness profiles to the lens galaxies in order to recover half-light radius values, allowing us to determine halo masses for lenses according to their size. Comparing our halo masses and sizes to baselines for that stellar mass yields a differential measurement of the halo mass-galaxy size relationship at fixed stellar mass, defined as $M_{h}(M_{*}) \propto r_{\mathrm{eff}}^{\eta}(M_{*})$. We find that on average, our lens galaxies have an $\eta = 0.42\pm0.12$, i.e. larger galaxies live in more massive dark matter haloes. The $\eta$ is strongest for high mass luminous red galaxies (LRGs). Investigation of this relationship in hydrodynamical simulations suggests that, at a fixed $M_{*}$, satellite galaxies have a larger $\eta$ and greater scatter in the $M_{\mathrm{h}}$ and $r_{\mathrm{eff}}$ relationship compared to central galaxies. 
\end{abstract}




\section{Introduction}
 
The relationship between baryonic matter and dark matter (DM) remains the source of many open questions in galaxy formation.  Studies of the mass-to-light ratio (\citealt{1979ARA&A..17..135F}) and star-to-halo mass ratio (SHMR) of galaxies both show that in general, more massive galaxies tend to live in more massive DM haloes. Techniques like abundance matching (\citealt{2002ApJ...569..101M}; \citealt{2004ApJ...609...35K}; \citealt{2006ApJ...647..201C}; \citealt{2006MNRAS.371.1173V})  show that this relationship is not simply linear or a power law, there is a deficiency of stars in low mass dwarf galaxies and high mass clusters. As well, there is a non-negligible amount of scatter (~0.2 dex) in these relationships (e.g. \citealt{2006MNRAS.371.1173V}; \citealt{2015MNRAS.454.1161Z}) with a currently unknown physical origin. It is therefore useful to search for 2nd order correlations between other parameters of the stellar distribution (such as galaxy size) and halo mass.

Several observational techniques have been used to estimate galaxy halo masses. Dynamical methods are commonly employed, but the need for tracers limits one to the region where the baryons dominate, while a full mass measurement requires data out to the virial radius. Using satellites or globular clusters as tracers can improve upon this, but they are not always available. 
In this paper, we will use weak gravitational lensing to measure halo masses directly at large radii where the halo is dominant over the baryons.

Galaxy-galaxy lensing (\citealt{1996ApJ...466..623B}; \citealt{1998ApJ...503..531H}) is the small distortion of background galaxies due to the lensing effect of a  foreground galaxy from that galaxy's baryons and dark matter halo, and, if that foreground galaxy is a satellite, the massive halo within which it resides. However, because the signal is very weak, it is necessary to stack many similar galaxies to obtain a significant measurement. From this one can infer a mass density distribution, and model the components of the galaxy and halo. This method allows observers to directly determine the average properties of the stacked lenses with considerable precision (\citealt{2006MNRAS.368..715M}; \citealt{2014MNRAS.437.2111V}; \citealt{2015MNRAS.447..298H}). Previous applications of weak lensing to the study of galaxy evolution have divided galaxies primarily by stellar mass, colour, and redshift. This simple subsampling allows investigation of the broad properties of dark matter as it relates to morphological type and stellar mass, but a more thorough investigation of the physics behind observed scaling relations or scatter, as we perform in this work, requires further subsampling. 

In this paper, we study the correlation between halo mass and galaxy size. For early-type galaxies, galaxy size has, for example, been linked to the dynamics of the inner regions of elliptical galaxies through the fundamental plane relationship. 
Size and environment are also related: the results of \cite{2007AJ....133.1741B} show that, for a given luminosity, brightest cluster galaxies (BCGs) are larger than non-BCGs. The properties of disc galaxies in regards to size are also not fully understood. Using the \cite{1977A&A....54..661T} relation and the baryonic Tully-Fisher relation (\citealt{1999AAS...19512901M}; \citealt{2000ApJ...533L..99M}) respectively, \cite{1999ApJ...513..561C} and \cite{2016ApJ...816L..14L} both observed that, at a fixed stellar mass, there was little to no size dependence on the rotational velocity.  

Size is a relatively simple morphological parameter to measure as long as a consistent definition is used. Automated surface brightness profile fitting has been shown to provide robust sizes for large samples of galaxies when care is taken to understand and control for the systematics that can influence the fits (\citealt{2007ApJS..172..615H}).

We will assume a power law relationship between galaxy size and halo mass at a fixed stellar mass $M_{h}(M_{*}) \propto r(M_{*})^{\eta}$. We fit this relationship separately for blue (primarily star-forming disc) and red (quiescent elliptical) galaxies in eight mass bins and three redshift bins. 

This paper is organized as follows. In Section \ref{sec:data} we give an overview of the CFHTLenS data used in our analysis. Section \ref{sec:Methods} describes how we find sizes and halo masses for our lens galaxies, and how we combine them to fit $\eta$. We present and discuss our observations in Section \ref{sec:results}. To aid in the analysis of our observational results, in Section \ref{sec:simcompare} we compare them with two hydrodynamical simulations. In Section \ref{sec:discussion} we primarily discuss physical interpretations for the observed $\eta$ values, the effects that assumptions we make have on our fits. Finally, we summarize our conclusions, and discuss approaches for expanding on our results in Section \ref{sec:conclusion}.

We adopt a flat $\Lambda$CDM cosmology, with Hubble parameter $H_{0} = 70$ km/s/Mpc, matter density parameter $\Omega_{\mathrm{m,0}} = 0.3$ and cosmological constant $\Omega_{\mathrm{\Lambda,0}} = 0.7$. All relevant quantities are derived using this value of $H_{0}$.

\section{Data}
\label{sec:data}

The data used in this paper are from the CFHT Lensing Survey \citep[hereafter CFHTLenS]{2012MNRAS.427..146H} analysis of the ``Wide'' portion of the CFHT Legacy Survey (CFHTLS). The imaging data used are comprised of the co-added science images described in \cite{2013MNRAS.433.2545E}. We use the CFHTLenS catalogues of photometric redshifts, luminosities and colours to sort our lens galaxies into bins, and fit surface brightness profiles to divide them into size bins. The catalogues also include shape measurements we use for galaxies photometrically defined as sources in our weak lensing analysis.  

\subsection{Imaging and photometry}

The CFHTLenS analysis is performed on the four CFHTLS wide fields (W1 through W4), imaging 154 square degrees. Imaging was performed in five bands, $u*$, $g'$, $r'$, $i'$, $z'$ to a magnitude limit of $i'_{AB} = 24.7$ using the MegaCam/MegaPrime wide-field imaging facility at the Canada France Hawaii Telescope (CFHT).

\subsection{Source galaxy ellipticities}
\label{sec:shapes}

We use the source galaxy ellipticities provided by CFHTLenS for $\mathrm{8.7\times10^6}$ background source galaxies (\citealt{2013MNRAS.429.2858M}). The source ellipticities have a Gaussian distribution with a scatter of $\sigma_{e} = 0.28$ (\citealt{2013MNRAS.432.2433H}). The uncertainty induced by this scatter is called ``shape noise''. Individual source shears are lie, on average, within an order of magnitude of ~0.01, which necessitates stacking many lenses. The source ellipticities are used as estimators for the gravitational shear induced by the lens, and when stacked allow for measurements of the average dark matter halo mass. The small additive correction to the source shapes applied in \cite{2012MNRAS.427..146H} and the multiplicative correction of \cite{2013MNRAS.429.2858M} are ignored here, as the statistical errors in our binned subsamples dominate over any small corrections.

\subsection{Redshifts and stellar masses}

Redshifts of lenses and sources in CFHTLenS, $z_{p}$, are determined photometrically as described in \cite{2012MNRAS.421.2355H}. The typical error in redshift increases from $\pm0.048$ for the closest lenses at $z_{p} = 0.2$ to $\pm0.092$ for the furthest sources at $z_{p} = 1.3$. 

The redshifts determined above are used to measure the stellar masses. The LePhare code, developed by \cite{2006A&A...457..841I} fits models of star formation history and dust extinction using the $u^*g'r'i'z'$ apparent magnitudes to find stellar masses and rest frame absolute magnitudes. \cite{2014MNRAS.437.2111V} describes in detail how the models of \cite{2003MNRAS.344.1000B} are used for these purposes. The spectral energy distribution templates used to fit the masses assume a \cite{2003PASP..115..763C} initial mass function, and exponentially decreasing star formation rates $\propto e^{-t/\tau}$ with nine different values for $\tau$ and two metalicities. 

\section{Methods and Measurements}
\label{sec:Methods}

\subsection{Samples}

Our sample consists of galaxies categorized as lenses or sources. The lenses are galaxies for which we measure sizes and that we stack to determine mean halo masses. For a given lens, the sources are background galaxies with lensing shape measurements that are averaged in radial bins and stacked. Lenses and sources are not mutually exclusive. The criteria outlined below permit a galaxy to be used as both a lens and a source in a stack, should they be met. 

\subsubsection{Lens galaxy sample}

Our lens sample consists of $2.06\times10^6$ galaxies with $i' < 23$, and $0.2 < z_{p} < 0.8$ from the CFHTLenS catalogues created with Source Extractor (SExtractor) running on the co-added science images. They are separated into red and blue samples based on their dust-corrected, rest frame $u^{*} - r'$ colours, dividing them at a value of 1.6, the location of the ``green valley'' of our lens sample. The redshift range is divided into three groups to account for possible evolution: $0.2 < z_{p} \le 0.4$, $0.4 < z_{p} \le 0.6$, and $0.6 < z_{p} \le 0.8$. 

We expect that $\eta$ may vary as a function of $M_{*}$ and by galaxy type, as different physical processes are relevant in low mass blue irregulars versus massive central cluster galaxies, for example. To account for this, we divide our lens sample into eight subsamples with different stellar masses. While we have stellar masses for individual galaxies, they are noisy, and using them to bin would introduce significant Eddington bias due to the slope of the mass function. Instead, we use the $r'$-band luminosities (which are less noisy and accordingly suffer from less bias) to place galaxies in bins such that when an appropriate stellar mass-to-light ratio is applied, the bins have a ~0.5 dex separation in $M_{*}$.

\subsubsection{Source galaxy sample}

The source sample consists of $5.6\times10^6$ galaxies that lie within at most one mask, with $i' < 24.7$ and $z_{p} < 1.3$, the upper limit for reliable photo-$z$ measurements in CFHTLenS. Galaxies may be used as both a lens or source as required if the criteria for a valid lens-source pair (below) is fulfilled. The magnification due to weak lensing is small, individually, and should have no measurable effect on the fitted galaxy size for lenses. The weights of each source account for measurement error and shape noise (\citealt{2013MNRAS.429.2858M}).

\subsubsection{Lens-source pairs}

The average redshift error for galaxies in the middle of our redshift range $z_{p} = 0.5$ is approximately 0.05. Therefore we select sources with a redshift separation $\Delta z_{p} > 0.1$, giving a $\sim 2\sigma$ buffer between pairs.

\subsection{Sizes of Lens Galaxies}
\label{sec:sizes}

In order to recover sizes for the lens galaxies in our sample, we use GALFITM (Bamford et al. in prep), a modification of GALFIT3 \citep{2002AJ....124..266P,2010AJ....139.2097P}. A postage-stamp cutout is created of each galaxy we wish to fit, with the dimensions determined by equations (2) and (3) in \cite{2007ApJS..172..615H}. The cutout size must be large enough to include as much of the light from the galaxy as possible, and to fit the sky background correctly. Larger cutouts also provide more sky pixels which can improve the fit, but the size must be limited to avoid the inclusion of too many contaminating stars and galaxies which must be masked out.

We fit a single S\'{e}rsic profile 
\begin{equation}
	\Sigma=\Sigma_{\mathrm{eff}}\exp{[\kappa((r/r_{\mathrm{eff}})^{\frac{1}{n}}-1)]}
	\label{eq:sersic}
\end{equation}
to the galaxies in the CFHTLenS catalogue, with $r_{\mathrm{eff}}$ (the half-light radius), $n$ (the S\'{e}rsic index), and $\Sigma_{\mathrm{eff}}$ (the brightness at $r_{\mathrm{eff}}$) as free parameters; $\kappa$ is dertermined by $n$ and is not free. We also fit its axis ratio, position angle and centroid coordinates. The sky background is fit with a central brightness and a linear gradient in the x and y axes across the image. GALFIT provides several options for light profiles, including double (bulge+disc) S\'{e}rsic fits. However since a significant number of the lenses are only just resolved, we prefer a more robust single S\'{e}rsic profile. Due to the number of galaxies fit we do not incorporate any deblending or simultaneous fitting for closely grouped galaxies. Handling of fits affected by this is discussed below. 

Given the large number of galaxies we wish to fit and the computation time limits, we did not deal with poor fits on an individual basis. Instead, we attempt to minimize the amount of fits that fail. A failed fit occurs when GALFITM crashes (very rare) or reaches its iteration limit yielding a flagged fit. Flagged fits contain parameters that are unlikely to be physical or reliable, such as an effective radius much smaller than a pixel, or very large S\'{e}rsic indexes. Putting a maximum and minimum size on our cutouts, and fitting the background separately, we lose <1\% of galaxies to flags. Flagged galaxies are not included in stacks, as their fitted sizes are unlikely to be reliable.

\subsection{Comparison with AEGIS}
\label{sec:quality}

In order to provide an objective check on the quality of our galaxy fits it is useful to compare similar fits to the same galaxies based on higher resolution space-based imaging. The AEGIS survey (\citealt{2007ApJ...660L...1D}) has imaged and catalogued many galaxies in the region around the Extended Groth Strip (EGS) using the Advanced Camera for Surveys (ACS) on the Hubble Space Telescope, and \cite{2012ApJS..200....9G} has fit S\'{e}rsic profiles to these AEGIS galaxies using GALFIT. The AEGIS region lies within four fields of the CFHTLS W3 patch. The primary differences between the two data sets are the PSF size (ACS: 0.1'', Megacam: varies 0.5''-1.0'') and the lack of a sky background in the fits to HST imaging.

Our lensing analysis is limited to lenses in the range of $0.2 < z_{p} < 0.8$ and $i' < 23$. We thus have approximately 3800 galaxies in the region covered by AEGIS for comparison. We expect that very small galaxies are likely to be smeared out to a larger apparent size by the larger PSF in CFHT imaging, comparing with ACS fits allows us to determine at which size this smearing becomes problematic. GALFIT convolves the S\'ersic profile with a provided PSF in fitting each galaxy, but suggests that unresolved sources ($r_{\mathrm{eff}}$ < 0.5 pixels) may be better fit with only the PSF. If many small galaxies have their sizes artificially increased, it would increase the median $r_{\mathrm{eff}}$ of the low-size bin used to fit $\eta$. This would reduce the $|\Delta r_{\mathrm{eff}}|$ without reducing $|\Delta M_{\mathrm{h}}|$ (described in Section \ref{sec:eta}), leading to a larger value of $\eta$ than with unaffected galaxies. 

The median fitted size of galaxies from CFHT is less than 0.2\% greater, on average, than ACS, while the fitted magnitudes are less than 0.02 brighter, indicating generally good agreement between S\'ersic profile fits, and thus $r_{\mathrm{eff}}$ for our ground-based imaging. The scatter in the fitted values of S\'ersic index, $r_{\mathrm{eff}}$, and apparent magnitude between he CFHT and ACS measurements increases at fainter magnitudes. Overall, there is a tendency for CFHT fits to have lower S\'ersic indexes ($\sim10$\% smaller) than the ACS fits. However this is not matched by an offset in magnitude, so while the shape of the surface brightness profile will be slightly impacted, the value of $r_{\mathrm{eff}}$ will not. The enlarging effect of atmospheric seeing is not significant. The only galaxies that have a significant systematic disagreement with AEGIS are so small as to be misidentified as stars by CFHT, preventing their inclusion in our sample. Detailed fit comparisons are included in Appendix \ref{sec:AEGIS}. 

\subsubsection{Purity of the CFHTLenS Galaxy Sample}

\begin{table}[!b]
	\caption{Number of EGS objects classified as galaxies or stars by SExtractor for CFHTLenS and AEGIS. Objects included in the final analysis are shown in bold.}
	\begin{tabularx}{\columnwidth}{l l l l l }
		& \multicolumn{2}{c}{Unflagged (3159)} &  \multicolumn{2}{c}{\hspace{8mm} Flagged (630)}\\
		\hline
		 & CFHT & CFHT & \hspace{8mm} CFHT & CFHT \\
		 &Galaxy & Star & \hspace{8mm} Galaxy &  Star\\
		\hline
		ACS Galaxy & \textbf{3037} & 59 &\hspace{8mm} 117 & 17\\
		ACS Star & \textbf{9} & 54 &\hspace{8mm} 12 & 479\\

	\end{tabularx}

	\label{tab:egs_objects}
\end{table}

Our initial galaxy sample consists of objects photometrically identified as galaxies from the CFHTLenS imaging in the SExtractor-created CFHTLenS catalogues; the final sample consists of objects from the initial sample that are successfully fit with an unflagged S\'{e}rsic profile. It is critical to minimize the contamination from stars in our galaxy samples because the stars have no lensing signal, and their inclusion in the averaging will reduce the excess surface density signal, thus leading to a lower halo mass when fitted. The AEGIS catalogues include their own star/galaxy classification based on HST observations with much better resolution which allows us to compare the relative ability to classify objects, with specific attention paid to objects classified differently by the two catalogues. The CFHTLS sample will not include galaxies that are misclassified by CFHT as stars, but retains stars misclassified by CFHT as galaxies.

We assume that for the $\sim3800$ objects included in both catalogues that the AEGIS classifications, being determined via space-based imaging, are correct. We allow GalfitM to attempt to fit a S\'{e}rsic profile to all objects including stars. As shown in Table~\ref{tab:egs_objects} GALFITM flags the fits of most stars, with only 9 out of the 3046 unflagged CFHT-classified galaxies being stars misidentified as galaxies. If we take this sample to be representative, then the expected stellar contamination is approximately 0.3\%. The systematic error due to this is small compared to the random errors.

\subsubsection{Completeness of galaxy sample}
\label{sec:completeness}

\begin{figure}
	\includegraphics[width=\columnwidth]{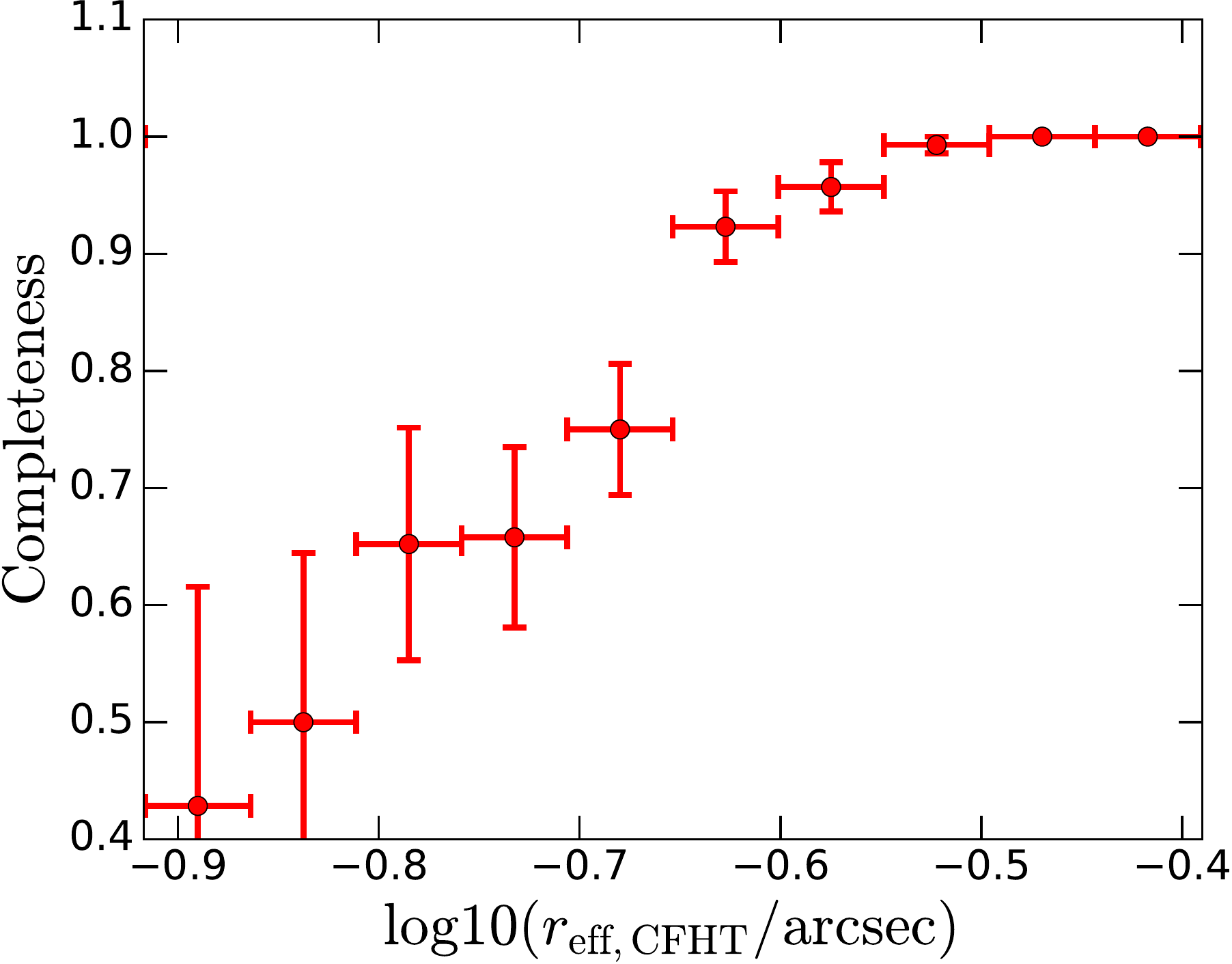}
	\caption{Completeness in bins of apparent size as determined from CFHT data. A value of 1 means that 100\% objects photometrically identified as galaxies by AEGIS are being identified as galaxies by CFHTLenS. A completeness of 0.5 means that 50\% of AEGIS galaxies are being identified correctly by CFHTLenS, with the remaining 50\% being misidentified as stars and not included in the final sample. The horizontal error bars indicate the width of the radial bins, and the vertical error bars are the $1\sigma$ uncertainty of the completeness using the Wald interval.}
	\label{fig:completeness}
\end{figure}

As shown in Table~\ref{tab:egs_objects} (see appendix~\ref{sec:AEGIS} for more information), there is some disagreement in star-galaxy separation for some objects in the AEGIS and CFHTLenS data sets, particularly at small apparent sizes where seeing heavily influences the resolution of CFHT imaging. The degree of disagreement in classification is shown in Figure~\ref{fig:completeness}. As the apparent size decreases, more galaxies begin to be incorrectly classified as stars, with 50\% of galaxies smaller than $0.15''$ being lost. The loss of small galaxies would shift the median galaxy size in all size subsamples to larger values. However, because our $\eta$ measurement is differential this amounts to a reduction in dynamic range.   

\subsection{Halo mass-size dependence $\eta$}
\label{sec:eta}

\begin{figure*}
	\includegraphics{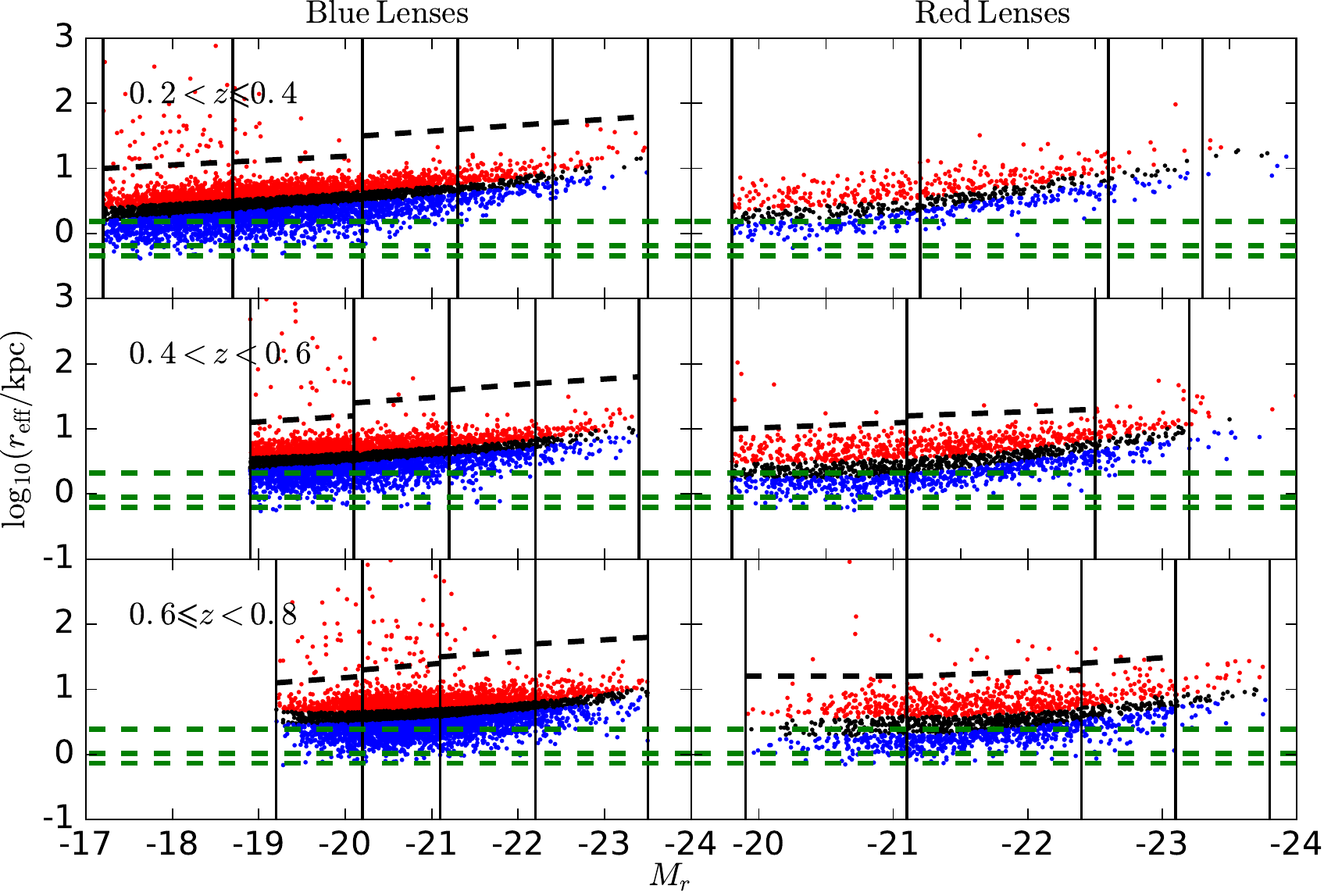}
	\caption{Luminosity-size relationship for a random subsample of the fitted blue (left) and red (right) lenses. Points are colour-coded red, black, and blue to denote their inclusion in the large, medium, and small stacks, respectively. The solid black lines mark the boundaries of each luminosity (or average stellar mass) bin. Galaxies larger than the dashed black lines are removed from the stacks, as they are unphysically large and usually the result of a poorly masked secondary object. The dashed green lines, from top to bottom indicate boundary of the 100\%, 50\%, and 0\% completeness regions (Section \ref{sec:completeness}).} 
	\label{fig:mag_radius}
\end{figure*}

We wish to make a differential measurement of the relationship between size and halo mass for a fixed luminosity, colour, and redshift bin. We assume a power-law relationship between halo mass $M_{\mathrm{h}}$ and half-light radius $r_{\mathrm{eff}}$ at fixed $M_{*}$ of the form:
\begin{equation}
	M_{\mathrm{h}}(M_{*}) \propto r_{\mathrm{eff}}^{\eta}(M_{*})
\end{equation}
In order to compare the relative strength of this relationship across all mass/luminosity bins, we compare $M_{h}$ and $r_{\mathrm{eff}}$ to fiducial masses and sizes for that stellar mass bin. For our fiducial mass, we use the expected stellar mass-to-halo mass ratio (SHMR) for a given $M_{*}$  (appendix C of \citealt{2015MNRAS.447..298H}). Our fiducial size is the median size of galaxies of a given $M_{*}$ from our fits. To account for differences between the expected halo mass of our average sized galaxies and observed galaxies we include a numerical factor $A$. Note that these expected masses are derived for central galaxies only, they allow us to make the differential mass measurements we require, but we do not expect them to match the average mass of stacked lenses which includes both central galaxies and satellites (i.e. A may not necessarily equal 1):
\begin{equation}
	\label{eqn:eta_rel}
	\frac{\langle M_{\mathrm{h}}\rangle}{M_{\mathrm{h,exp}}(M_{*})} = A \left( \frac{ r_{\mathrm{eff,med}}}{r_{\mathrm{eff,med}}(M_{*})} \right) ^{\eta}
\end{equation}  
Since we stack our lenses, we use $\langle M_{\mathrm{h}}\rangle$ and $r_{\mathrm{eff, med}}$, which, respectively, are the ensemble average halo mass, and the median size of the lenses in a given size bin, whereas $M_{\mathrm{h,exp}}(M_{*})$ and $r_{\mathrm{eff,med}}(M_{*})$ are, respectively, the expected halo mass and median size for galaxies of all sizes in a given stellar mass bin.

We perform our analysis in $\log-\log$ space, so for simplicity, we introduce
\begin{equation}
\Delta r'_{\mathrm{eff}}(M_{*})= \log_{10}\left(\frac{r_{\mathrm{eff, med}}}{r_{\mathrm{eff,med}}(M_{*})}\right)
\end{equation}
and
\begin{equation}
\Delta M'_{\mathrm{h}}(M_{*})= \log_{10}\left(\frac{\langle M_{\mathrm{h}}\rangle}{M_{\mathrm{h,exp}}(M_{*})}\right)
\end{equation}

Figure~\ref{fig:mag_radius} shows our $r_{\mathrm{eff}}$ vs $L$ relationship and demonstrates how we adaptively split our galaxies into small, average and large size bins for their mass. Galaxies that have extremely large $r_{\mathrm{eff}}$ are likely to have been poorly fit due to inadequately masked bright companions, bright sky background, and so on. These galaxies are removed and the lenses in each size bin are stacked and fitted with a halo model as described in Section~\ref{ref:models}.

\subsection{Halo masses of lens galaxies}
\label{ref:models}

We determine the average dark matter halo mass for galaxies in each of our three size subsamples at a fixed stellar mass. This cannot be done on an individual basis due to measurement and statistical errors so we must first stack the measured tangential shears for each lens source pair as a function of radius. These stacked shears are then used to fit a dark matter profile, giving us the mass we require for our analysis.
 
\subsubsection{Average shear}

To determine halo masses from weak lensing, we need to determine the excess surface mass density
\begin{equation}
	\Delta\Sigma(R)=\overline{\Sigma(<R)}-\overline{\Sigma(R)}
\end{equation}
which is the difference between the projected average surface mass within a circle of radius $R$ and the surface density at that radius. The tangential shear $\gamma_{t}$ can be estimated by averaging the ellipticities of background source galaxies along an axis that is perpendicular to a line connecting the lens and and that source. This tangential shear is directly related to the excess surface density through the equation
\begin{equation}
	\Delta\Sigma(R)=\Sigma_{\mathrm{crit}}\langle\gamma_{t}(R)\rangle
\end{equation} 
where $\Sigma_{\mathrm{crit}}$ is the critical surface density, used to define the Einstein radius of the lens. It is given by
\begin{equation}
	\Sigma_{\mathrm{crit}}=\frac{c^2}{4\pi G}\frac{D_{\mathrm{s}}}{D_{\mathrm{l}} D_{\mathrm{ls}}}
\end{equation}
where $D_\mathrm{s}$, $D_\mathrm{l}$, and $D_{\mathrm{ls}}$ are the angular diameter distances to the source, lens, and lens-source distance respectively.

The shape noise is combined with the measurement error on the ellipticities within \textit{lens}fit to give a weight $w$ which is applied when stacking (described in \citealt{2013MNRAS.429.2858M}). Each pair is also weighted by $W=\Sigma_{\mathrm{crit}}^{-2}$ following \cite{2015MNRAS.447..298H}. When stacked, the average excess surface density of our lenses is
\begin{equation}
	\langle\Delta\Sigma(R)\rangle=\frac{\sum w_{j} \gamma_{t,j}\Sigma_{\mathrm{crit},ij}W_{ij}}{\sum w_j W_{ij}}
\end{equation}
which is summed over all lenses $i$, and sources $j$ in a given radial bin.  

\subsubsection{Halo Model}
\label{sec:halomodel}

\begin{figure*}
	\includegraphics[width=\textwidth]{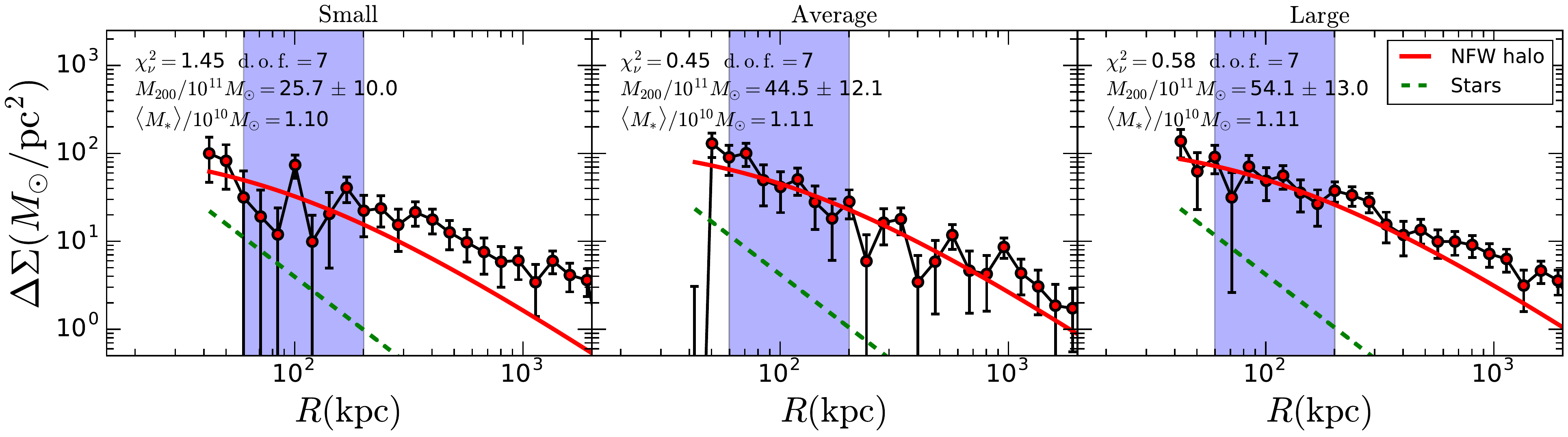}
	\caption{Example stacked $\Delta\Sigma$ data from CFHTLenS red galaxies of $\langle M_{*} \rangle = 1.2\times10^{11}M_{\odot}$ and $\langle z_{\mathrm{p}} \rangle = 0.5$ in three size bins, from left to right: small, average, and large. The blue shaded region contains the points used for the one-halo NFW fit, according to Section \ref{sec:halomodel}. The solid red line is the one-halo NFW term and the green dashed line is the stellar component.}
	\label{fig:nfw_examples}
\end{figure*}

The data are fit with a simplified version of the halo model described in Section 3 of \cite{2015MNRAS.447..298H}. We focus on the one-halo term of this model, describing the contribution from the stacked lenses' own stars and dark matter. We model $\Delta\Sigma(R)$ far from the stars, so we can treat the stellar mass as a point mass:

\begin{equation}
	\Delta\Sigma_{*}(R) = \frac{\langle M_{*}\rangle}{\pi R^2}
	\label{eq:ds_star}
\end{equation}

The dark matter halo is modeled by an NFW (\citealt{1997ApJ...490..493N}) density profile and is parameterized by its virial mass $M_{200}$. The halo's concentration $c_{200}$, and thus scale radius $r_{s}$, is fixed by its redshift $z$ and $M_{200}$ using the relaxed halo model from \cite{2011MNRAS.411..584M}. The projected halo surface mass excess term $\Delta\Sigma_{\mathrm{NFW}}$ is described in \cite{1996ApJ...466..623B} \& \cite{1996A&A...313..697B}. The full simplified one-halo term is:

\begin{equation}
	\Delta\Sigma_{\mathrm{1h}}(R) = \Delta\Sigma_{\mathrm{NFW}}(R) + \Delta\Sigma_{*}(R)
\end{equation}

Unlike in \cite{2015MNRAS.447..298H} we do not fit the offset group halo term. This term is a convolution of all of the central haloes that satellite lenses are embedded in. The offset group halo term dominates at intermediate radii, outside of the region in which the one-halo term is relevant. Correctly fitting the offset-group halo term requires information about the environments in which the lens galaxies reside. The approach used in \cite{2015MNRAS.447..298H} assumes a halo occupation distribution (HOD) using the method of \cite{2012A&A...542A...5C} to recover a satellite fraction $f_{\mathrm{sat}}$, the expected fraction of galaxies of a given mass that reside in the halo of a larger group or galaxy. Using the same method here would be assuming that $r_{\mathrm{eff}}$ is not affected by environment. For example, if tidal stripping is responsible for $r_{\mathrm{eff}}$ differences, more stripped galaxies should reside closer to the centers of clusters. We elect not to fit the offset group term as doing so requires assumptions about the satellite fraction and radial distribution of satellites for subsamples of each size (but see Appendix \ref{sec:offset} for a discussion of these assumptions). For similar reasons, we do not truncate the one-halo term (\citealt{2009JCAP...01..015B}), which arises for satellite galaxies due to stripping and their existence within a more massive halo.

We have restricted the radial range of our fits to radii within which the offset-group term is sub-dominant to the one-halo term (see Figures 2 and 3 in \citealt{2015MNRAS.447..298H}) and also wish to avoid extended light from the foreground galaxy affecting the quality of source shape measurements (\citealt{1998ApJ...503..531H};\citealt{2011MNRAS.412.2665V}). The radial range is set using the fits from \citealt{2015MNRAS.447..298H} as a guideline; the inner (stellar) boundary is set by the point where the NFW term is ten times greater than the stellar term, and the outer (group) boundary is set by where the NFW term is twice the offset-group term. Figure \ref{fig:nfw_examples} shows an example of the three halo masses we fit for each size bin, highlighting in blue the range over which we fit, for a particular $M_{*}$ and $z_p$ bin. In some cases, the fitting regions may contain points with only negative $\Delta\Sigma$ values due to noisy source shapes. In these instances, we fit $\eta$ using the remaining two size bins.  

For each mass-colour-redshift bin we use equation \ref{eqn:eta_rel} to fit the relationship between $\Delta M'_{\mathrm{h}}(M_{*})$ and $\Delta r'_{\mathrm{eff}}(M_{*})$ using our three size subsamples to find $\eta$. The intercept $A$ should be 1 if the mass of our average size subsample matches our model, but this comes from the fits to central galaxies only. Our $M_{*}$-colour bins contain both centrals and satellites with possible differences in $f_{\mathrm{sat}}$ between size bins. $A\neq1$ simply indicates that galaxies at the median size of a given stellar mass are not all centrals. Thus, to determine $\eta(M_{*})$ we fit: 
\begin{equation}
	\Delta M'_{\mathrm{h}}(M_{*}) = \eta(M_{*}) \Delta r'_{\mathrm{eff}}(M_{*})  + \log_{10}(A)
\end{equation}

\section{Results}
\label{sec:results}

\begin{figure}
	\includegraphics[width=\columnwidth]{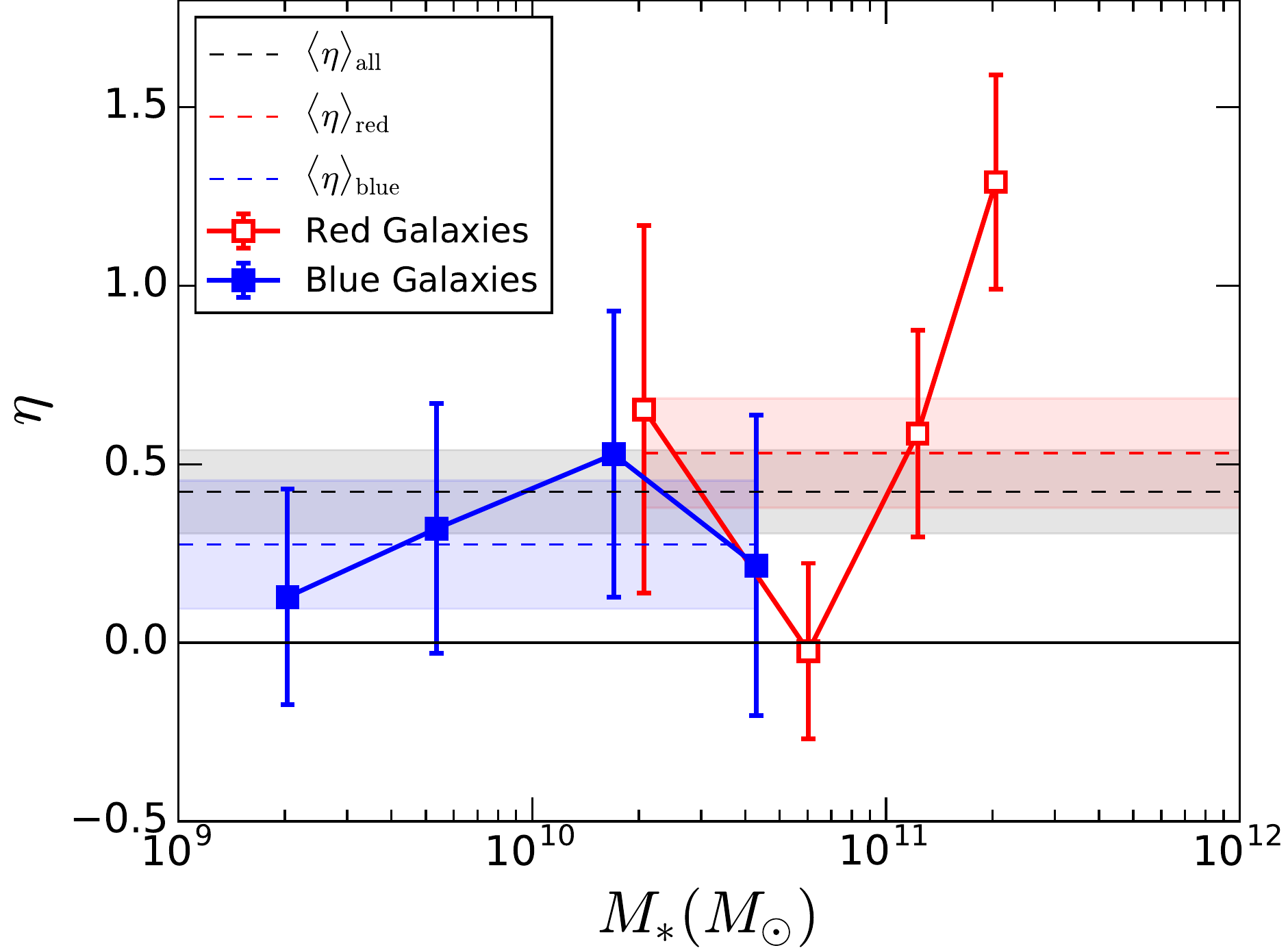}
	\caption{Average halo mass-size dependence for our red and blue galaxy samples. The fiducial average slope of all bins is plotted as the dashed black line, with the grey band representing the $1\sigma$ region. The average $\eta$ and $1\sigma$ regions of red and blue galaxies is plotted with red and blue dashed lines and shaded regions, respectively. The points represent the weighted averages of the $\eta$ values and average $M_{*}$ of the galaxies each $(z,M_{*})$ bin.}
	\label{fig:allslopes}
\end{figure}

\begin{table*}
	\centering
	\caption{Halo mass-size fit results for our stacked galaxy bins. On the left of the vertical line are the details for each of our mass/redshift/colour bins, to the right of the line are the same bins, averaged over redshift. N indicates the number if lenses stacked in each size bin.}
	\begin{tabularx}{1.0\textwidth}{lllllll|lll} 
		Colour & $z_{p}$ range & $N_{\mathrm{small}}$ & $N_{\mathrm{medium}}$ & $N_{\mathrm{large}}$ & $\log_{10}(\langle M_{*}\rangle)$ & $\eta$ & $\log_{10}(\langle M_{*}\rangle_{z})$ & $\langle \eta \rangle$ & $\sigma$\\
		\hline
		\hline
		Blue & $\mathrm{0.2 < z_{p} \leqslant 0.4}$ & 73207 & 74588 & 77050 & $\mathrm{9.231}$ & $\mathrm{0.17\pm 0.34}$ & $$ & $$ & $$\\
		Blue & $\mathrm{0.4 < z_{p} < 0.6}$ & 66846 & 72654 & 74662 & $\mathrm{9.279}$ & $\mathrm{0.00\pm 0.77}$ & $\mathrm{9.308}$ & $\mathrm{0.13\pm 0.30}$ & $\mathrm{0.43}$\\
		Blue & $\mathrm{0.6 \leqslant z_{p} < 0.8}$ & 31028 & 31030 & 32013 & $\mathrm{9.398}$ & $\mathrm{-0.1\pm 1.2}$ & $$ & $$ & $$\\
		\hline
		Blue & $\mathrm{0.2 < z_{p} \leqslant 0.4}$ & 27156 & 27626 & 28565 & $\mathrm{9.708}$ & $\mathrm{0.67\pm 0.52}$ & $$ & $$ & $$\\
		Blue & $\mathrm{0.4 < z_{p} < 0.6}$ & 46276 & 46270 & 47680 & $\mathrm{9.732}$ & $\mathrm{0.44\pm 0.53}$ & $\mathrm{9.730}$ & $\mathrm{0.32\pm 0.35}$ & $\mathrm{0.91}$\\
		Blue & $\mathrm{0.6 \leqslant z_{p} < 0.8}$ & 87781 & 91681 & 94500 & $\mathrm{9.748}$ & $\mathrm{-1.6\pm 1.1}$ & $$ & $$ & $$\\
		\hline
		Blue & $\mathrm{0.2 < z_{p} \leqslant 0.4}$ & 10660 & 10848 & 11331 & $\mathrm{10.241}$ & $\mathrm{0.75\pm 0.55}$ & $$ & $$ & $$\\
		Blue & $\mathrm{0.4 < z_{p} < 0.6}$  & 18852 & 20752 & 21425 & $\mathrm{10.220}$ & $\mathrm{-0.03\pm 0.75}$ & $\mathrm{10.230}$ & $\mathrm{0.53\pm 0.40}$ & $\mathrm{1.33}$\\
		Blue & $\mathrm{0.6 \leqslant z_{p} < 0.8}$ & 58980 & 62987 & 64760 & $\mathrm{10.231}$ & $\mathrm{0.78\pm 0.93}$ & $$ & $$\\
		\hline
		Blue & $\mathrm{0.2 < z_{p} \leqslant 0.4}$ &1750 & 1805 & 1909 & $\mathrm{10.620}$ & $\mathrm{-0.38\pm 0.64}$ & $$ & $$ & $$\\
		Blue & $\mathrm{0.4 < z_{p} < 0.6}$ & 5112 & 5561 & 5737 & $\mathrm{10.620}$ & $\mathrm{0.43\pm 0.85}$ & $\mathrm{10.634}$ & $\mathrm{0.22\pm 0.42}$ & $\mathrm{0.52}$\\
		Blue & $\mathrm{0.6 \leqslant z_{p} < 0.8}$ & 14553 & 15509 & 15933 & $\mathrm{10.660}$ & $\mathrm{0.85\pm 0.75}$ & $$ & $$ & $$\\
		\hline
		\hline
		Red & $\mathrm{0.2 < z_{p} \leqslant 0.4}$ & 12966 & 13313 & 13965 & $\mathrm{10.330}$ & $\mathrm{0.46\pm 0.71}$ & $$ & $$ & $$\\
		Red & $\mathrm{0.4 < z_{p} < 0.6}$ & 23397 & 24876 & 25565 & $\mathrm{10.301}$ & $\mathrm{0.87\pm 0.75}$ & $\mathrm{10.316}$ & $\mathrm{0.65\pm 0.52}$ & $\mathrm{1.25}$\\
		\hline
		Red & $\mathrm{0.2 < z_{p} \leqslant 0.4}$ & 11529 & 11772 & 12325 & $\mathrm{10.790}$ & $\mathrm{-0.02\pm 0.32}$ & $$ & $$ & $$\\
		Red & $\mathrm{0.4 < z_{p} < 0.6}$ & 28079 & 30314 & 30857 & $\mathrm{10.770}$ & $\mathrm{0.13\pm 0.40}$ & $\mathrm{10.780}$ & $\mathrm{-0.02\pm 0.25}$ & $\mathrm{0.08}$\\
		Red & $\mathrm{0.6 \leqslant z_{p} < 0.8}$ & 36633 & 38588 & 39499 & $\mathrm{10.780}$ & $\mathrm{-2.2\pm 1.5}$ & $$ & $$ & $$\\
		\hline
		Red & $\mathrm{0.2 < z_{p} \leqslant 0.4}$ & 1589 & 1620 & 1685 & $\mathrm{11.100}$ & $\mathrm{0.25\pm 0.46}$ & $$ & $$ & $$\\
		Red & $\mathrm{0.4 < z_{p} < 0.6}$ & 4728 & 5095 & 5248 & $\mathrm{11.077}$ & $\mathrm{0.59\pm 0.51}$ & $\mathrm{11.089}$ & $\mathrm{0.59\pm 0.29}$ & $\mathrm{2.03}$\\
		Red & $\mathrm{0.6 \leqslant z_{p} < 0.8}$ & 9014 & 9536 & 9807 & $\mathrm{11.090}$ & $\mathrm{1.04\pm 0.61}$ & $$ & $$ & $$\\
		\hline
		Red & $\mathrm{0.2 < z_{p} \leqslant 0.4}$ & 269 & 272 & 285 & $\mathrm{11.290}$ & $\mathrm{1.13\pm 0.53}$ & $$ & $$ & $$\\
		Red & $\mathrm{0.4 < z_{p} < 0.6}$ & 822 & 870 & 921 & $\mathrm{11.310}$ & $\mathrm{1.40\pm 0.39}$ & $\mathrm{11.310}$ & $\mathrm{1.29\pm 0.30}$ & $\mathrm{4.3}$\\
		Red & $\mathrm{0.6 \leqslant z_{p} < 0.8}$ & 1967 & 2079 & 2154 & $\mathrm{11.330}$ & $\mathrm{1.15\pm 0.94}$ & $$ & $$ & $$\\
	\end{tabularx}
	\label{tab:full_results}
\end{table*}

The fitted values of $\eta$ are shown in Figure \ref{fig:allslopes} and Table \ref{tab:full_results}. We do not find any significant differences in $\eta$ at different redshifts, so for each mass bin we combine our three $z_{p}$ bins to create an average $\eta$ weighted by the uncertainty of each $z_{\mathrm{p}}$ bin's $\eta$ fit. Within $1\sigma$, all of our stellar mass bins show a positive correlation between size and halo mass.

Given the uncertainties in our $\eta$ values, there does not appear to be a significant trend in $\eta$ with stellar mass. If we simply average all mass and colour bins, we find $\langle\eta\rangle=0.42\pm0.12$. The $\langle\eta\rangle$ of blue galaxies is $0.28\pm0.18$ and the $\langle\eta\rangle$ of red galaxies is $0.53\pm0.15$. This indicates that, on average, there is a relationship between halo mass and size at fixed $M_{*}$. However, this correlation is weak, and can not account fully for the scatter in the SHMR ($\sim0.2$ dex). The scatter in log size $\sigma (\Delta r'_{\mathrm{eff}})$ is approximately 0.15 dex, giving
\begin{equation}
	\sigma (\Delta M'_{\mathrm{h}}) \approx \langle\eta\rangle \, \sigma (\Delta r'_{\mathrm{eff}}) \sim 0.06 \, \textrm{dex} ,
\end{equation}
just over $30\%$ of the observed scatter.

In our fits, the NFW halo model has a concentration $c_{200}$ which is not a free parameter, but is determined by the lens' $z_{\mathrm{p}}$ and $M_{h}$ (see Section~\ref{sec:halomodel}).  The effect of changing $c_{200}$ at fixed $M_{\mathrm{h}}$ is illustrated in Figure~\ref{fig:conctest}. At a certain radius, $r_{\mathrm{eq}}$, all concentrations will have the same $\Delta\Sigma$ value ($\sim100-300\mathrm{kpc}$ over our $M_{*}$ range). More (less) concentrated haloes have more (less) mass contained at $r \ll r_{\mathrm{eq}}$, which is capable of mimicking a more (less) massive halo if the fit depends primarily upon radial bins within $r_{\mathrm{eq}}$.  

One might hope to measure the concentration as well as the halo mass using the detailed shape of  $\Delta\Sigma(R)$. In practice, however this is difficult given the uncertainty of fitting, for example, the offset group term simultaneously.  With realistic weak lensing data, there is therefore, in practice, an approximate  degeneracy between $M_{h}$ and $c_{200}$. For example, \cite{2013MNRAS.431.1439G} illustrates that a halo of $\log_{10}(M_{\mathrm{h}}) = 11.6$ with $c_{200}=8$ predicts roughly the same lensing signal as a halo of $\log_{10}(M_{\mathrm{h}}) = 11.7$ with $c_{200}=4$. Consequently, one can imagine an alternative scheme where we hold $M_{\mathrm{h}}$ fixed and instead fit for $c_{200}$, and then plot $\Delta c'_{200}$ as a function of $\Delta r'_{\mathrm{eff}}$, yielding $\eta_{c}$.  Thus, in general, our results might be interpreted as a change in concentration with galaxy size, with larger galaxies corresponding to a higher halo concentration.

\section{Comparison with hydrodynamical simulations}
\label{sec:simcompare}

\begin{figure*}
	\includegraphics[width=\textwidth]{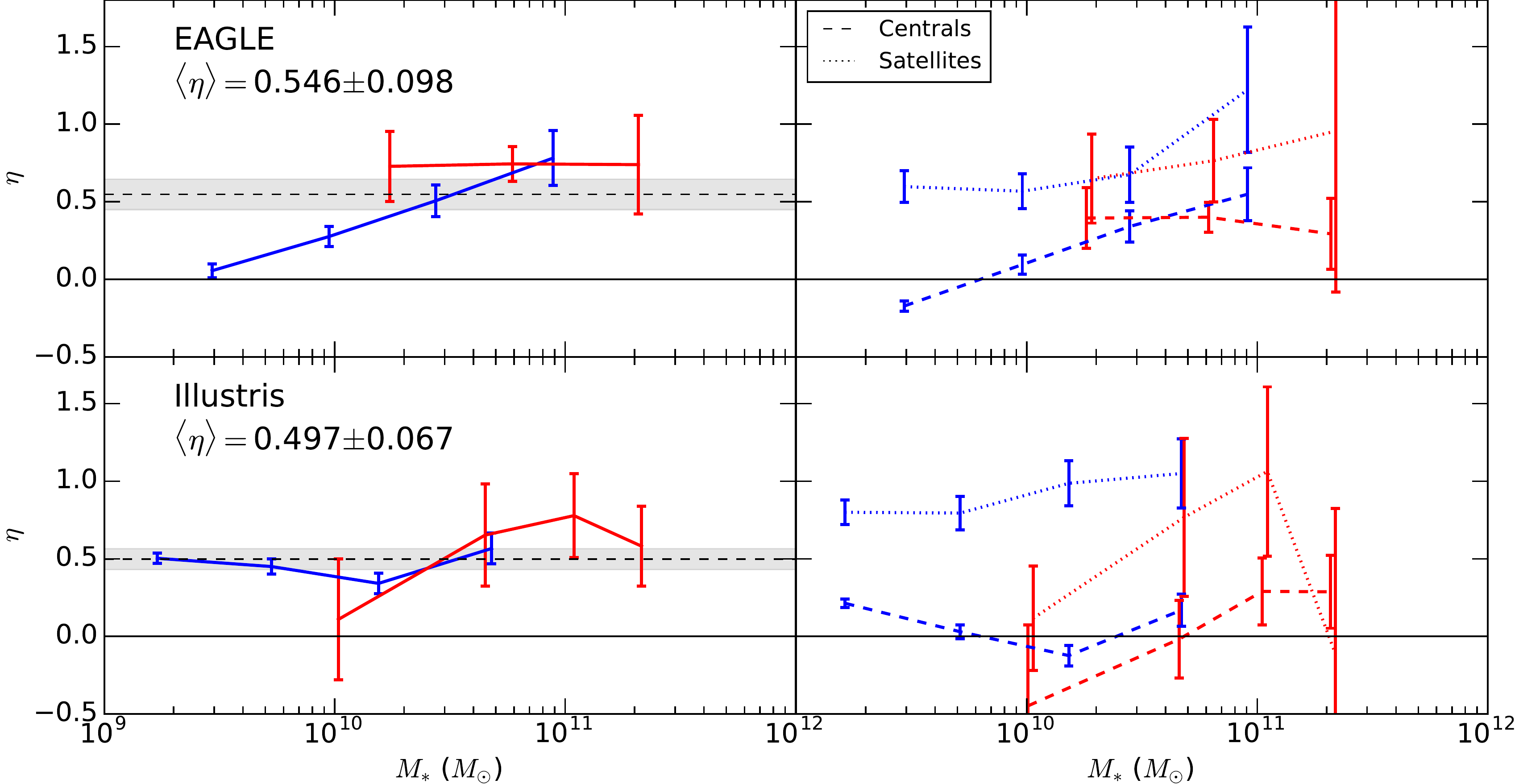}
	\caption{Halo mass-size dependence for the simulations examined. Top row: EAGLE, bottom row: Illustris. Left: Mass bins without splitting by environment, includes fiducial average. Right: The same mass bins split into satellite galaxies and central galaxies, dashed lines represent centrals, dotted lines represent satellites. The red central masses have been shifted 5\% lower for clarity.}
	\label{fig:simulation_slopes}
\end{figure*}

Due to the reliance of our weak lensing measurements upon stacking, we cannot examine $\eta$ using individual lenses, and cannot measure the scatter around the trends that we fit. Our data also do not include environmental information, and one important factor we wish to investigate is whether $\eta$ differs for central galaxies versus those in subhaloes.

We investigate two hydrodynamical simulations, EAGLE (\citealt{2015MNRAS.446..521S}; \citealt{2015MNRAS.450.1937C}) and Illustris (\citealt{2014MNRAS.444.1518V}), which allows us to separately examine the size-halo mass trend for centrals and subhaloes as well as determine what information is inaccessible due to the stacking and averaging required in order to perform our lensing analysis. Using the simulations, we wish to create a set of sample galaxy bins with broadly the same characteristics as our data. Each simulation has a snapshot at $z = 0.5$, the average redshift of our galaxy sample. We use the same method of splitting galaxies into subsamples by size as in our observations. We take the mean $M_{\mathrm{h}}$ of all galaxies in each size bin and compare that to $M_{\mathrm{h,exp}}(\langle M_{*} \rangle)$.

Galaxies are divided by colour in a manner similar to the data; each simulation includes stellar luminosities in the SDSS $ugriz$ filters. We look for bimodality in the colour distributions, as we do with our data, in order to separate the simulated galaxies into red and blue samples. It is worth noting that the luminosities in our catalogue are corrected for dust extinction, while the simulated galaxies' magnitudes explicitly exclude dust. Slight differences in the mass-to-light ratio lead to differences in the average stellar mass of galaxies in each bin across the simulations and data, but they broadly cover the same ranges.

Rather than use a half-light radius from simulated observations, we use the half-stellar mass radius for our sizes. EAGLE's catalogues include both the 3D and projected half-mass radius. We use the projected radii for a more direct comparison to our observations. Illustris includes only the 3D half-mass radius. Investigating the difference using the EAGLE galaxies shows that the 3D radii are on average $30\%$ larger than the projected radius, with no change as a function of $M_{*}$. 
Since our $\eta$ relies on a differential measurement of size, we consider Illustris' $r_{\mathrm{eff,M_{*},3D}}$ acceptable for our purposes under the assumption that the same relationship holds. 
Note that Illustris is also known to be in disagreement with several observed properties of galaxy populations: the luminosity-size relationship is too shallow at high masses, and flattens at low masses; there is also a lack of low-mass bulge galaxies (\citealt{2017arXiv170108206B}). The fits to the simulated galaxies are summarized in Figure~\ref{fig:simulation_slopes}; values and fits for each stellar mass bin are tabulated and shown in appendix~\ref{sec:simdata}. 

\subsection{Simulation Results}
\begin{figure}
	\includegraphics[width=\columnwidth]{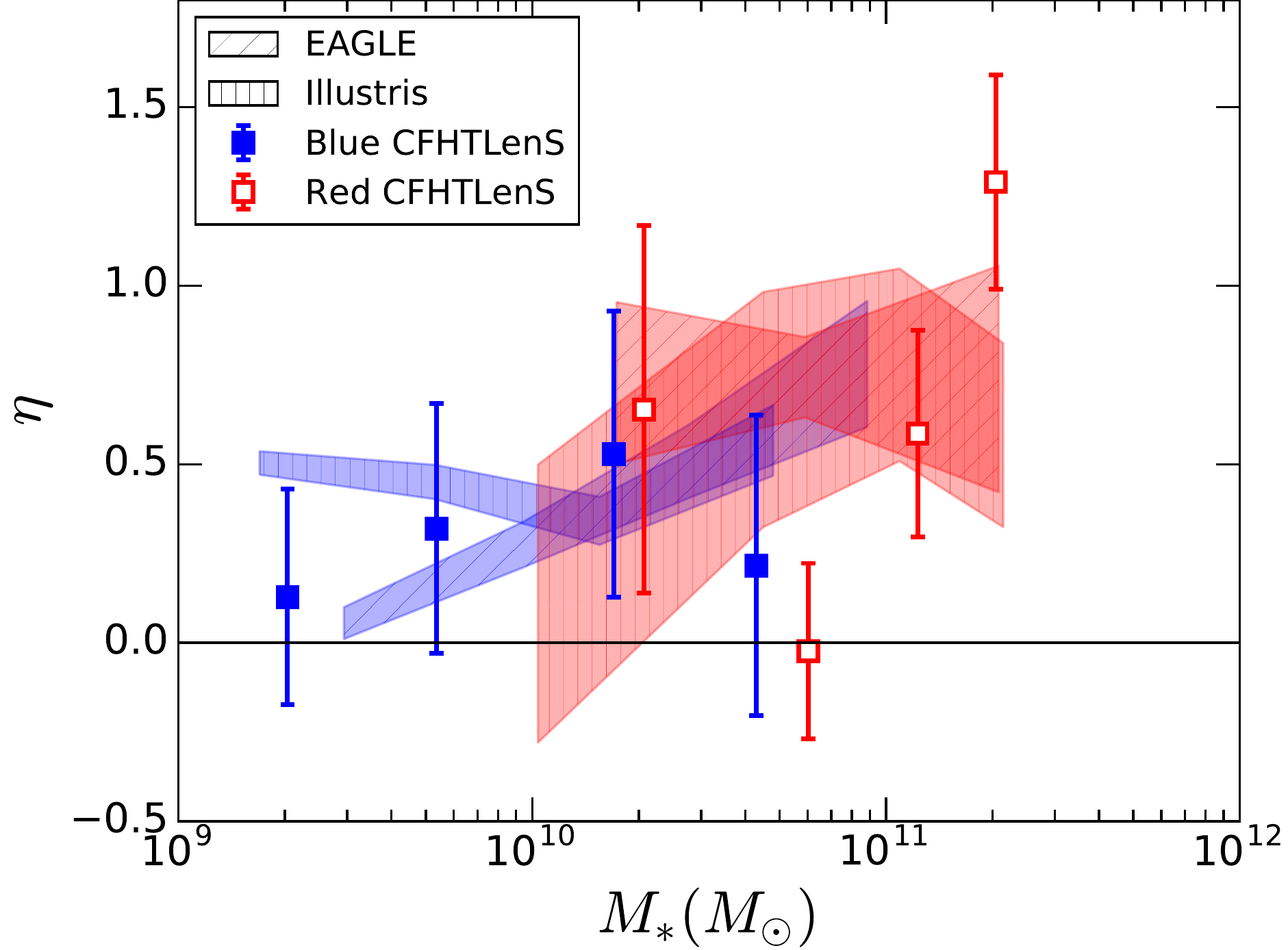}
	\caption{Halo mass-size dependence for EAGLE's and Illustris' red and blue galaxy samples without splitting by environment. Our data is overlaid on the range covered by the simulation slopes + errors (shaded regions). Filled symbols are blue galaxies and open symbols are red galaxies. The simulations and our data agree within $1\sigma$.}
	\label{fig:simslopes}
\end{figure}

Figure~\ref{fig:simslopes} shows our fitted $\eta$ values from the CFHTLenS overlaid on the range of slopes covered by the two hydrodynamical simulations we examined. Our data lie within $1\sigma$ of the range spanned by the simulations. We can draw several clear conclusions from the simulations:
\begin{enumerate}
\item Satellite galaxies universally show a stronger $\Delta M'_{\mathrm{h}}$-$\Delta r'_{\mathrm{eff}}$ relationship than centrals. $\eta_{\mathrm{sat}}$ lies in the range of 0.5-1.0 while $\eta_{\mathrm{cent}}$ is generally between 0-0.5.
\item The simulations show that the scatter in the $\Delta M'_{\mathrm{h}}$-$\Delta r'_{\mathrm{eff}}$ relationship is much tighter for central galaxies than for satellites. A likely reason for this is the large variety of processes capable of influencing each parameter in the cluster environment when compared to galaxies in the field or at the cluster center, such as tidal stripping, ram pressure stripping, and harassment.
\item In most $M_{*}$ bins, central/field galaxies dominate the population. However, bins with higher satellite fractions do not have significantly higher $\eta$. This is likely due to the more tightly correlated central halo relationship dominating the fit. 
\item In general, the $\eta$ of the simulations are broadly consistent with each other. 
\item Where the simulations do differ is in low mass red and blue galaxies. At low mass, Illustris blue galaxies show a a greater $\eta$ than EAGLE, while Illustris red galaxies show a lower $\eta$ than EAGLE.
\end{enumerate}

\section{Discussion}
\label{sec:discussion}
We have found that $\eta$ is generally positive, and this result is also seen in the EAGLE and Illustris simulations. In this section, we consider several physical effects that may be responsible for this. To model how halo mass and size may be related, we examine how specific processes could change the position of a given galaxy on a plot of $\Delta M'_{\mathrm{h}}$ vs $\Delta r'_{\mathrm{eff}}$. We consider both effects due to initial conditions and in situ evolution, and environmental effects, due to interactions with other galaxies or due to the cluster environment. From the simulations we expect that satellite and central galaxies will have different $\eta$.

\subsection{Halo and Galaxy Properties}

For central galaxies, the properties of the DM halo may drive properties of the galaxy that ultimately forms within it. In addition to halo mass, two such properties are concentration and angular momentum.

\subsubsection{Concentration}
The link between halo concentration and halo mass accretion rate in N-body simulations is shown in e.g. \cite{2002ApJ...568...52W}; for haloes of a fixed total mass at the present time, older, faster-forming haloes have higher concentrations. \cite{2017MNRAS.465.2381M} argue that concentration explains much of the scatter in $M_{*}$ for a fixed $M_{h}$ in the EAGLE simulations. Specifically, at fixed halo mass, a larger stellar mass corresponds to a more concentrated halo. To explain our results however, this must also be tied to size. It is not clear how the size of the galaxy might be correlated with the concentration of the halo in which it forms.  Naively, one might expect that a more concentrated halo would lead to a more concentrated galaxy. This is clearly an interesting area for future study.

\begin{figure}
	\includegraphics[width=\columnwidth]{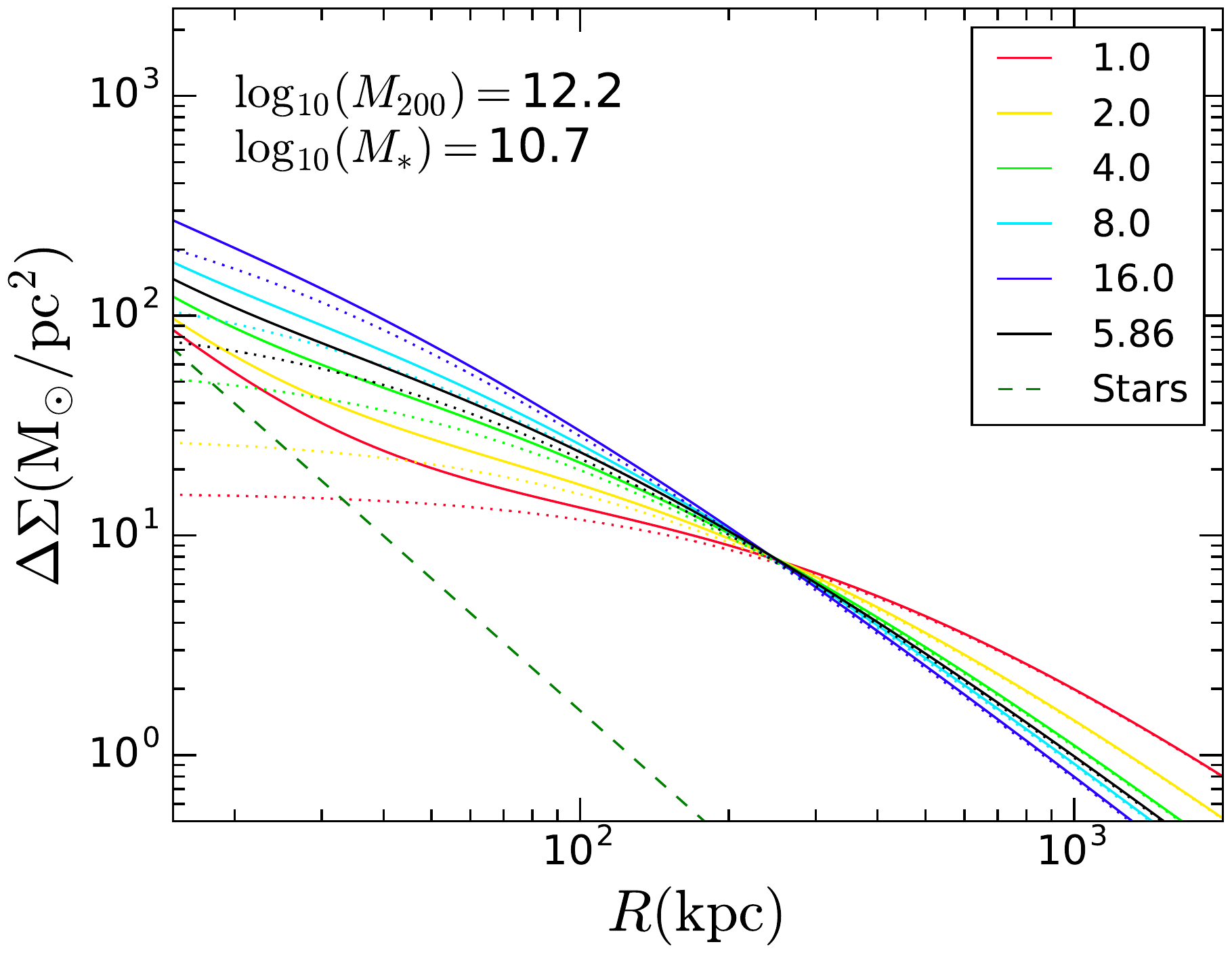}
	\caption{Excess surface density plot illustrating how $c_{200}$ can appear to shift the one-halo term's normalization at small radii. $M_{h}$ and $M_{*}$ remain fixed. The colours indicate different values of $c_{200}$. The radius at which all profiles meet is $r_{\mathrm{eq}}$. The dotted lines are the NFW halos alone, and the solid lines are the combined NFW + stars. The black lines indicate the profile for the expected halo concentration of a galaxy of this mass at $z_{\mathrm{p}}=0.5$. Note that $c_{200}\lesssim 4$ are not realistic and are included for illustration only.}
	\label{fig:conctest}
\end{figure}

\subsubsection{Angular momentum}

It is thought that tidal torques on the baryon/dark matter proto-halo create a spin (\citealt{1969ApJ...155..393P}), which is observed in N-body simulations to have a spread \citep{1987ApJ...319..575B}. In the model of \cite{1980MNRAS.193..189F}, this angular momentum is conserved as the disc shrinks and so a disc's final rotationally-supported radius is related to the primordial spin of its halo. For two galaxies of equal $M_{\mathrm{h}}$ and equal amounts of star-forming gas mass, the halo with more angular momentum will form a larger disc with a lower gas surface density and, by the Kennicutt-Schmidt relationship (\citealt{1959ApJ...129..243S}; \citealt{1998ApJ...498..541K}), might be expected, over cosmic history, to form fewer stars. This would manifest as a positive $\eta$.  

Observationally, however, there is little evidence of any correlation between the circular velocity, a proxy for halo mass, and disc size for isolated (or low density environment) disc galaxies \citep{1999ApJ...513..561C, 2016ApJ...816L..14L}. This result is consistent with the prediction, found above, for central blue galaxies in Illustris.  In contrast, in EAGLE there is a correlation for central discs.

\subsection{Mergers}
The role of galaxy mergers in the halo mass-size relationship should be treated with consideration of several factors: whether the galaxy lives in a high density or low density environment, is a disc or elliptical-type, and mass ratios of past mergers. In general we expect mergers will increase both the size and mass of the resultant galaxy, but the amount can vary based on the above factors. To first order, galaxies that have gone through more mergers should be larger and have more massive haloes and stellar components. In order to account for the observed halo mass-size relationship, we must consider how to create a range of halo masses and sizes while keeping stellar mass relatively fixed. 

The bin with the largest $\eta$ is our highest $M_{*}$ red bin which contains mostly LRGs. Evidence suggests that minor mergers (along with in situ star formation) account for the majority of the present day mass growth of central cluster/group galaxies with major mergers only becoming important at the present day (\citealt{2017arXiv170109012G}). \cite{2009ApJ...699L.178N} describes the rate of size growth of central elliptical galaxies due to minor mergers. As mergers occur, stellar mass is distributed outward leading to size growth. The equation describing the growth in size after accretion is:
\begin{equation}
\label{eq:sizegrowth}
	\frac{\langle r_{f}\rangle}{\langle r_{i}\rangle} = \frac{(1+\lambda)^2}{(1+\lambda\epsilon)}
\end{equation}
where $\lambda$ is the ratio between the initial stellar mass and the accreted stellar mass and $\epsilon$ is the ratio of mean squared speeds of the accreted material to the initial material (this is \textasciitilde1 for major mergers, and \textasciitilde0 for minor mergers). As a baseline, we consider the scenario in which an LRG doubles its stellar mass through mergers; if all mergers are minor mergers this leads to a factor of four increase in size and if this occurs through major mergers the galaxy doubles in size. This first order approximation assumes that all stars and dark matter are accreted and that the galaxy has time to virialize between each merger.

We need to know the growth of both $r_{\mathrm{eff}}$ and $M_{\mathrm{h}}$ at a fixed $M_{*}$:
\begin{equation}
	\Delta \log_{10}\left(r_{\mathrm{eff}}\right)|_{M_{*}} = \Delta \log_{10}\left(r_{\mathrm{eff}}\right) - \alpha \Delta \log_{10}\left(M_{*}\right)
\end{equation}
$\alpha$ for red galaxies is taken from our $r_{\mathrm{eff}}-M_{*}$ fits (see appendix \ref{sec:AEGIS}). Following equation \ref{eq:sizegrowth} for minor mergers:
\begin{equation}
	\Delta \log_{10}\left(r_{\mathrm{eff}}\right)|_{M_{*}} = \log_{10}\left(\frac{4 r_{\mathrm{eff,i}}}{r_{\mathrm{eff,i}}}\right) - 0.29 \log_{10}\left(\frac{2 M_{*\mathrm{,i}}}{M_{*\mathrm{,i}}}\right)
\end{equation}
and for halo mass:
\begin{equation}
\label{eq:massgrowth}
	\Delta \log_{10}\left(M_{\mathrm{h}}\right)|_{M_{*}} = \Delta \log_{10}\left(M_{\mathrm{h}}\right) - s_{\mathrm{h}}\left(M_{*}\right) \Delta \log_{10}\left(M_{*}\right)
\end{equation}
where $s_{\mathrm{h}}(M_{*})$ is the slope of the $M_{*}-M_{\mathrm{h}}$ relationship at fixed $M_{*}$. This comes from the model used to determine our expected $M_{\mathrm{h}}$ discussed in Section \ref{sec:eta}. The halo mass growth depends on the assumptions made for the SHMR of accreted galaxies. A simple approximation is to assume that on average, we can use the universal fraction of baryons in stars:
\begin{equation}
\gamma = \frac{\Omega_{*}}{\Omega_{m}}=\frac{0.0018}{0.30}
\end{equation}
where we use the value of $\Omega_{*}$ as determined by GAMA (\citealt{2016MNRAS.457.1308M}). Thus the growth of $M_{\mathrm{h}}$ is
\begin{equation}
	M_{\mathrm{h},f} = \left(\frac{1}{\gamma}+\frac{1}{f(M_{*})}\right)M_{*}
\end{equation}
Where $f(M_{*,i})$ is the SHMR at the initial $M_{*}$. The change in $M_{\mathrm{h}}$ is
\begin{equation}
\Delta \log_{10}\left(M_{\mathrm{h}}\right) = \log_{10}\left(1+\frac{f}{\gamma}\right)
\end{equation}
and equation \ref{eq:massgrowth} becomes
\begin{equation}
	\Delta \log_{10}\left(M_{\mathrm{h}}\right)|_{M_{*}} = \log_{10}\left(1+\frac{f}{\gamma}\right) - s_{\mathrm{h}}\left(M_{*}\right) \log_{10}\left(\frac{2M_{*,i}}{M_{*,i}}\right)
\end{equation}
For a galaxy in our highest $M_{*}$ bin, this leads to $\eta \approx 1$, within $1\sigma$ of the $\eta$ we observe for LRGs. 

The above is a simple case where all accreted galaxies are close to the universal baryon ratio. In a more realistic model, the mass distribution of mergers experienced by the galaxy are likely to follow a \cite{1974ApJ...187..425P} mass function truncated at the original galaxy's mass (e.g. \citealt{2004MNRAS.355..819G}; \citealt{2009MNRAS.398.1858M}). Since our simplified model assumes all mergers are minor, a more realistic model will have more mergers at a mass ratio closer to 1, thus we expect the true $\eta$ should be lower.

Another simplification we made is to assume that mergers happen instantaneously. The stellar mass will take time to merge, so $r_{\mathrm{eff}}$ will not immediately quadruple while they remain separated. The haloes of the infalling satellites will be stripped first and join the halo of the central galaxy. What this leads to is a greater than expected $M_{\mathrm{h}}$ while $r_{\mathrm{eff}}$ remains at closer to the original value. Due to this, we can consider the $\eta$ in our toy model to be a lower limit.  

\subsection{Stripping}
Stripping is predicted to play a large role in the $\eta$s values that we observe for satellites. The high density of galaxies in clusters provide ample opportunities for tidal stripping, while at the same time galaxies experience ram-pressure stripping due to the intracluster medium. These two stripping processes are likely to occur simultaneously, making analytical modeling difficult. We examine them in isolation, taking this into consideration for any conclusions we draw. 

Tidal stripping is the removal of material due to close interactions with other galaxies in the cluster, with the least bound matter stripped off first. Since dark matter haloes appear to be much greater in extent than the stellar portions of galaxies, for the first few Gyr within the cluster, tidal interactions preferentially strip the dark matter halo (\citealt{2013MNRAS.431.3533C}; \citealt{2016ApJ...833..109S}) and some gas (\citealt{2000ApJ...540..113B}) while leaving the stars relatively intact. Tidal stripping effectively truncates the dark matter halo and reduces its mass. Eventually, continued tidal stripping removes significant amounts of both stars and dark matter, and galaxies will continue to evolve to lower halo masses, moving to smaller sizes as well. In the absence of mergers and accretion this could lead to galaxies moving to different stellar mass bins.

The most straightforward way that tidal stripping alone would affect galaxies in our $M_{\mathrm{h}}$-$r_{\mathrm{eff}}$ plots would be to, over time, move a galaxy to lower halo masses. For a satellite of a given $M_{*}$, the larger $r_{\mathrm{eff}}$ is, the less its halo can be stripped before removing stellar mass, so smaller satellites are able to have more of their halo stripped, reducing their halo mass in comparison to their larger counterparts. This effect would correspond to a positive $\eta$.    

A galaxy moving through the intracluster medium will experience a ram pressure that depends on the density of the medium and its velocity, and that will preferentially strip star forming gas (\citealt{1972ApJ...176....1G}). The ram pressure first strips the lower density gas in the outer regions of the galaxy, depositing it into a wake. This leads to reduced star formation within the galaxy, and, over time, a reduced size (\citealt{2009A&A...499...87K}). Dark matter is unaffected by ram pressure, however two stripping mechanisms could work in tandem; reducing $M_{\mathrm{h}}$ through tides, and $r_{\mathrm{eff}}$ through ram pressure to create a positive $\eta$.

One consideration in attempting to explain the observed $\eta$ through stripping is that more-stripped lenses are more likely to be satellites, and so there would be a stronger offset group term contribution to the observed $\Delta\Sigma$ in smaller size bins. This is considered in more detail in Appendix \ref{sec:offset}.  There we show that this effect would tend to make $\eta$ larger than the values quoted above, possibly by as much 0.2-0.3 for bins with high satellite fractions, i.e.\ red satellites with $M_{*} \lesssim ??$

\section{Conclusion}
\label{sec:conclusion}

We have obtained sizes for more than $2\times10^6$ lens galaxies, and used $5.6\times10^6$ source galaxy shape measurements from CFHTLenS to determine the relationship between a galaxy's size and its dark matter halo mass at a fixed $M_{*}$ for galaxies between $10^9 M_{\odot}$ and $3\times10^{11}\odot$. This is the first work to provide observational evidence for a correlation of this nature. The relationship takes the form of $M_{\mathrm{h}}\propto r_{\mathrm{eff}}^{\eta}$.  We compared these observations to the hydrodynamical simulations, EAGLE and Illustris. We divided a sample of red and blue simulated galaxies at $z_{\mathrm{p}}$ = 0.5 into four $M_{*}$ bins similar to our observational sample. We use the half mass radius to substitute for the efective radius and the total dark matter mass  of the associated particles as a substitute for the lensing mass. We also observe a positive $\eta$ in the simulations. The conclusions we draw are:
\begin{enumerate}
\item The weighted average $\eta$ across all $M_{*}$ bins is $0.42\pm0.16$ indicating a positive correlation between size and halo mass.
\item We do not detect a strong evolutionary trend across our three $z_{p}$ bins centered at $0.2$, $0.4$, and $0.6$.
\item The value of $\eta$ is highest in our highest $M_{*}$ bin, which contains LRGs, but the uncertainties on our other $\eta$ measurements do not permit us to make a strong statement regarding trends as a function of stellar mass.
\item We show that a toy minor merger model can explain the high $\eta$ observed for LRGs.
\item Our measurements of $\eta$ are consistent with the predictions from hydrodynamical simulations for all but the most massive galaxies. The limited size of the simulation boxes (~100Mpc) allows only very few high mass galaxies to form, limiting the accuracy of any fits to their $\Delta M'_{\mathrm{h}}$-$\Delta r'_{\mathrm{eff}}$ relationship.
\item The simulations indicate that the $\eta$ for satellites is significantly higher than for central galaxies. They also show that the scatter in the $\Delta M'_{\mathrm{h}}$-$\Delta r'_{\mathrm{eff}}$ relationship is much tighter for centrals than for satellites. This indicates that there are likely more environmental processes driving $\eta$ for satellite galaxies, such as tidal stripping affecting the stars and dark matter.
\end{enumerate}

The role of environmental effects on the halo mass-size relationship is apparent in the simulations we investigated. Some form of environmental information would be instrumental in future studies primarily to separate lenses by environment to compare the strength of the relationship between satellite, central, and field subsamples. 
  
\section*{Acknowledgments}

We acknowledge the substantial efforts of both the CFHT staff in implementing the Legacy Survey, and of the CFHTLenS team in preparing catalogues of galaxy ellipticities and photometric redshifts.

This work is based on observations obtained with MegaPrime/MegaCam, a joint project of CFHT and CEA/IRFU, at the Canada-France-Hawaii Telescope (CFHT) which is operated by the National Research Council (NRC) of Canada, the Institut National des Sciences de l'Univers of the Centre National de la Recherche Scientifique (CNRS) of France, and the University of Hawaii. This research used the facilities of the Canadian Astronomy Data Centre operated by the National Research Council of Canada with the support of the Canadian Space Agency. CFHTLenS data processing was made possible thanks to significant computing support from the NSERC Research Tools and Instruments grant program.

MJH and MLB acknowledge support from NSERC.




\bibliographystyle{mnras}
\bibliography{paper_bib} 




\appendix
\section{Comparison With AEGIS}
\label{sec:AEGIS}

\begin{figure*}
	\includegraphics[width=\textwidth]{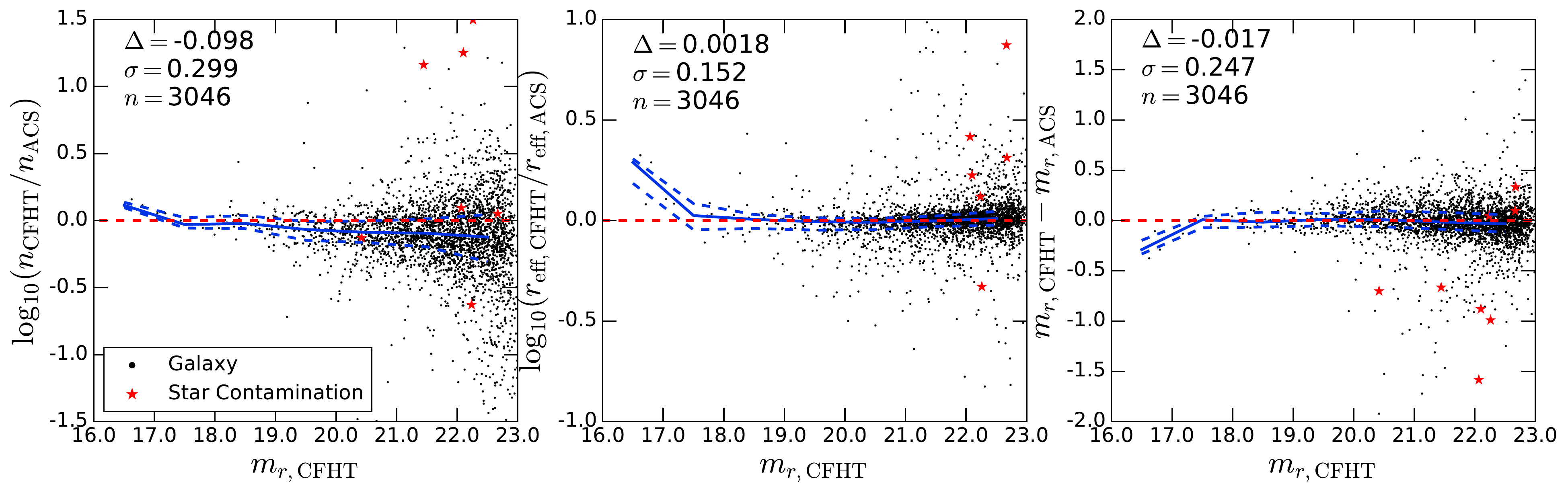}
	\caption[width=\paperwidth]{Comparison of GALFIT results for galaxies in common to AEGIS and CFHTLenS. Black points are objects classified as galaxies by SExtractor in both surveys and red points are objects classified as stars in AEGIS which in CFHTLS are mis-classified as galaxies. The blue line is the median offset in each parameter, the dashed blue line indicates the semi-interquartile range.}
	\label{fig:fit_grid}
\end{figure*}

\begin{figure*}
	\includegraphics[width=\textwidth]{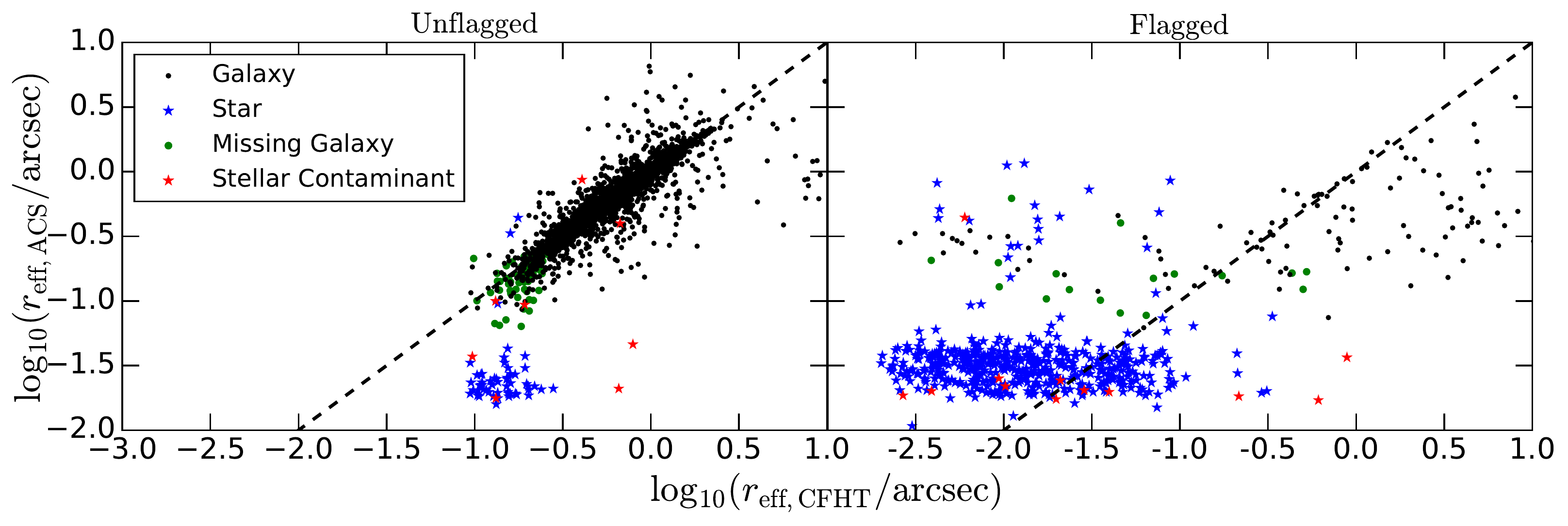}
	\caption{Apparent size comparison of the objects in the AEGIS overlap. The left panel shows objects with unflagged model fits, the right panel shows objects with flagged model fits. Missing galaxies have been identified as galaxies in the higher resolution HST imaging, and identified as stars in the CFHT imaging, so they will not be included in our stacks. The stellar contamination is caused by stars that are misidentified as galaxies in the CFHTLenS catalogues. The unflagged objects show no systematic disagreements in apparent size at small radii, showing that GALFIT reliably models apparent size for galaxies which it can fit successfully.} 
	\label{fig:apparent_size}
\end{figure*}

\begin{figure}
	\includegraphics[width=\columnwidth]{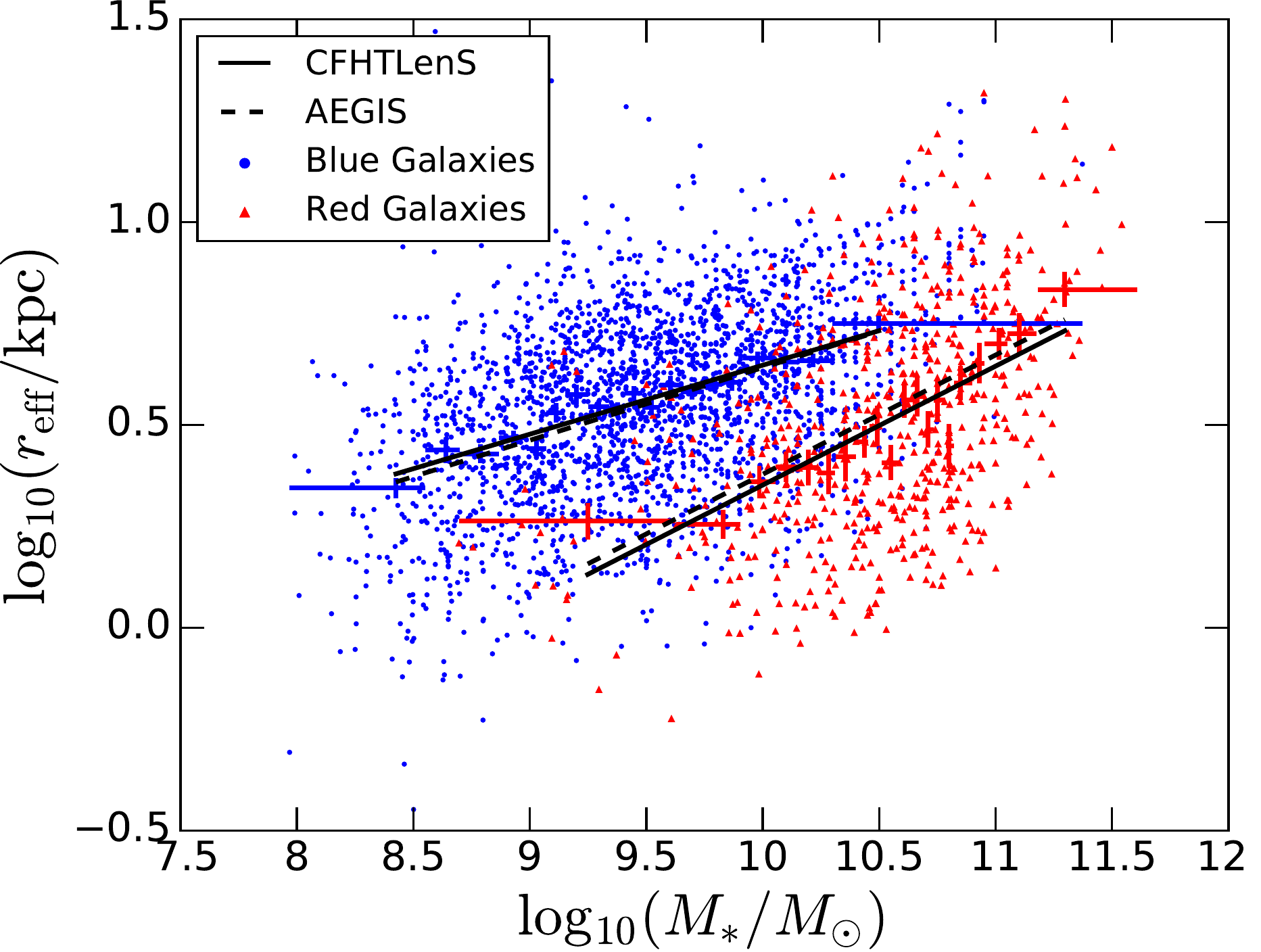}
	\caption{Mass-Size relationship of EGS galaxies by CFHTLS. Points are size fits as determined by this work. Crosses represent the median mass and size of the galaxies in a each bin. The solid black lines are size-mass fits to the CFHTLenS data, the dashed lines are the fits to the AEGIS data.  }
	\label{fig:mass_radius}
\end{figure}

\begin{table}[!b]
	\caption{Power-law fits for the mass-size relationship of the AEGIS overlap galaxies.}
	\begin{tabularx}{\columnwidth}{c c c}
		Galaxy Sample & $\alpha$ & $\beta$ \\
		\hline
		CFHT Blue & \hspace{5mm} 0.170 $\pm$ 0.009 & \hspace{5mm} -1.05 $\pm$ 0.08 \\
		ACS Blue & \hspace{5mm} 0.179 $\pm$ 0.009 & \hspace{5mm} -1.15 $\pm$ 0.09 \\
		CFHT Red & \hspace{5mm} 0.29 $\pm$ 0.02 & \hspace{5mm} -2.6 $\pm$ 0.2 \\
		ACS Red & \hspace{5mm} 0.29 $\pm$ 0.02 & \hspace{5mm} -2.6 $\pm$ 0.3 \\
	\end{tabularx}
	\label{tab:mass_size_fits}
\end{table}

The galaxy-by-galaxy comparison of the three main S\'ersic parameters as a function of magnitude is shown in Figure~\ref{fig:fit_grid}. The average scatter in the effective radius fits is ~0.15 dex and does not change significantly over the magnitude range of interest. When the apparent sizes of CFHT galaxies are very small, they are impacted by atmospheric seeing, smearing them out and making them appear larger they are. This is seen in Figure~\ref{fig:apparent_size} comparing the apparent sizes of objects as determined by ACS and CFHT. The smallest galaxies marked in green show a consistent tendency to appear larger to CFHT, but they are misidentified as stars and prevented from inclusion in the lens stacks. The unflagged, correctly identified galaxies show no such offset, and so we can be confident that within our magnitude limits, we can accurately recover galaxy sizes via GALFIT.

As a final check to the quality of our size fits relative to AEGIS, we fit the $M_{*}$-$r_{\mathrm{eff}}$ relationship for the CFHT and ACS fits and compare. These fits are shown in  Figure~\ref{fig:mass_radius}. We bin galaxies into groups of 120 for blue galaxies, and 30 for red galaxies, as there are roughly four times more blue galaxies in our sample. Next we fit a power law to the median $r_{\mathrm{eff}}$ of each bin as a function of $M_{*}$ of the form

\begin{equation}
	\log_{10}(r_{\mathrm{eff,med}}) = \alpha \cdot \log_{10}(M_{*,\mathrm{med}}) + \beta
	\label{eq:mass_size_log}
\end{equation}
\begin{equation}
	r_{\mathrm{eff,med}} = M_{*,\mathrm{med}}^{\alpha} \cdot 10^{\beta}
	\label{eq:mass_size_pwr}
\end{equation}

The results of the fits are shown in Table~\ref{tab:mass_size_fits}, and we see no significant systematic difference between AEGIS fits and CFHT fits in the $M_{*}$-$r_{\mathrm{eff}}$ relationship. This indicates that while fits to individual galaxies may differ, no significant systematic differences are present that, on average, would result in an improper division of galaxies into our size bins.

\section{Lensing Results}
\label{sec:lensingresults}

\begin{figure*}
	\subfloat{\includegraphics[width=\textwidth]{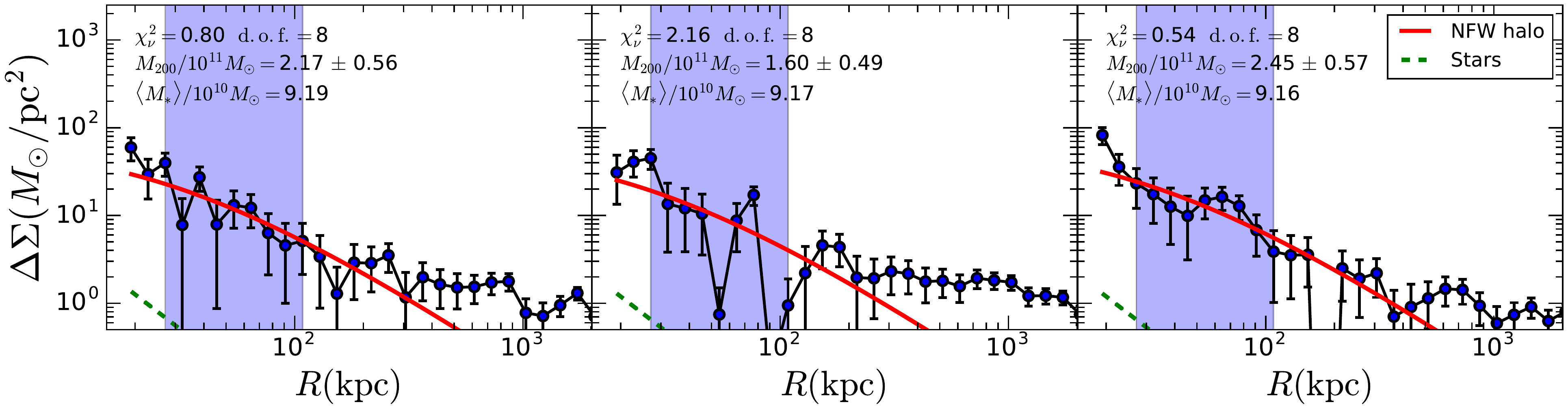}}\newline
	\subfloat{\includegraphics[width=\textwidth]{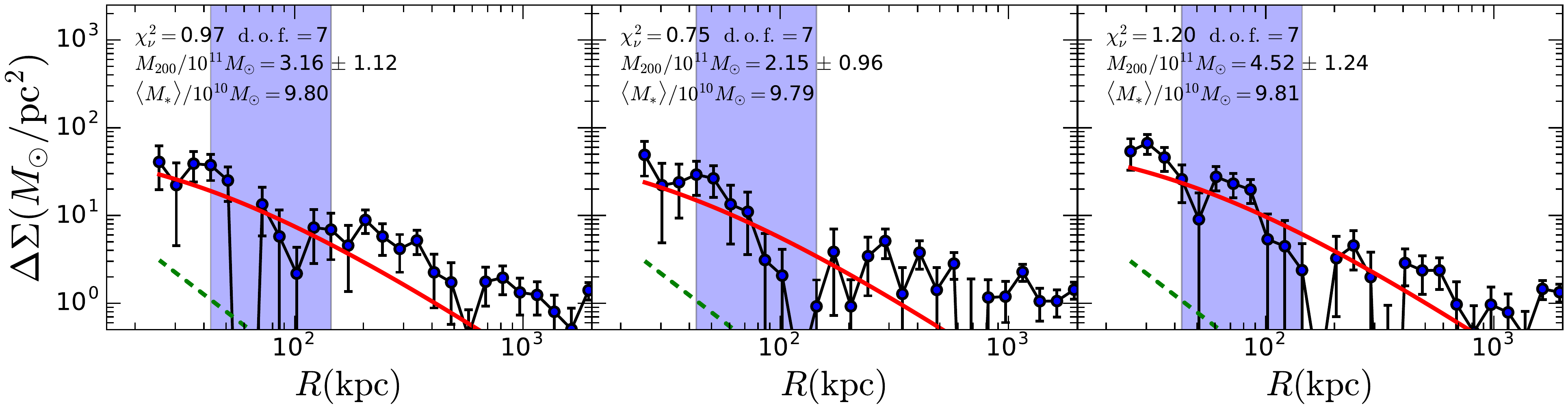}}\newline
	\subfloat{\includegraphics[width=\textwidth]{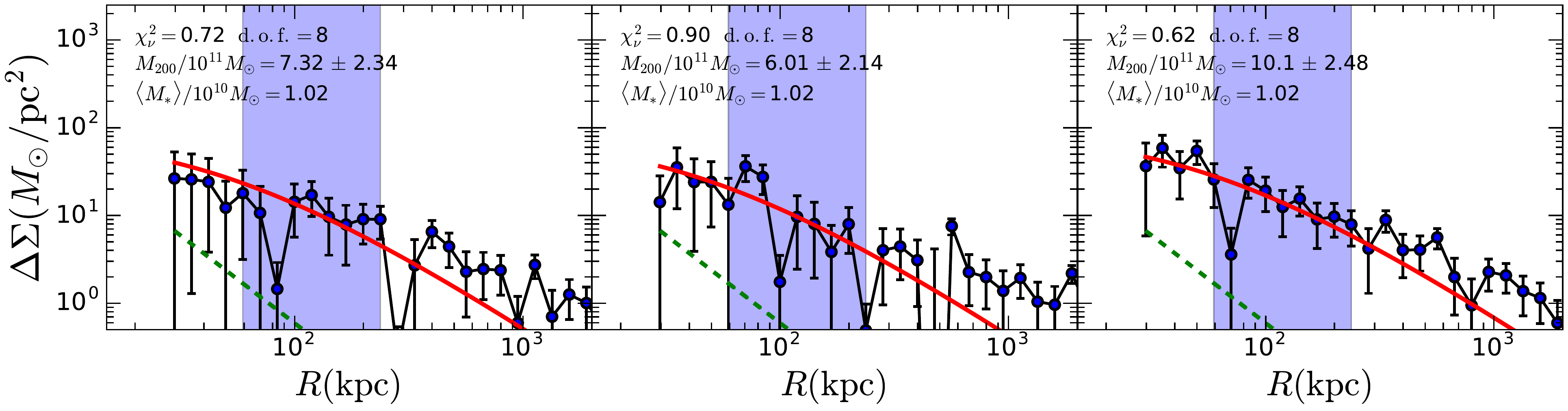}}\newline
	\subfloat{\includegraphics[width=\textwidth]{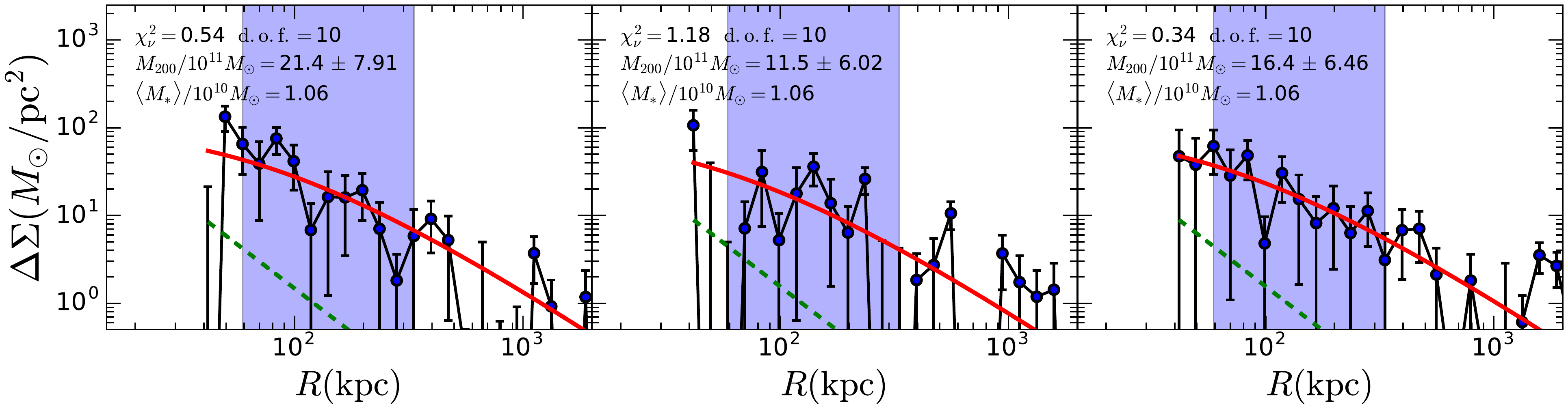}}
	\caption[Blue low-z NFW fits]{Stacked $\Delta\Sigma$ data from CFHTLenS blue galaxies, $0.2<z<0.4$, in three size bins, from left to right: small, average, and large. The blue shaded region contains the points used for the one-halo NFW fit, according to Section \ref{sec:halomodel}. The solid red line is the one-halo NFW term and the green dashed line is the stellar component. The average lens mass increases from top to bottom}
	\label{fig:nfw_b_z03}
\end{figure*}

\begin{figure*}
	\subfloat{\includegraphics[width=\textwidth]{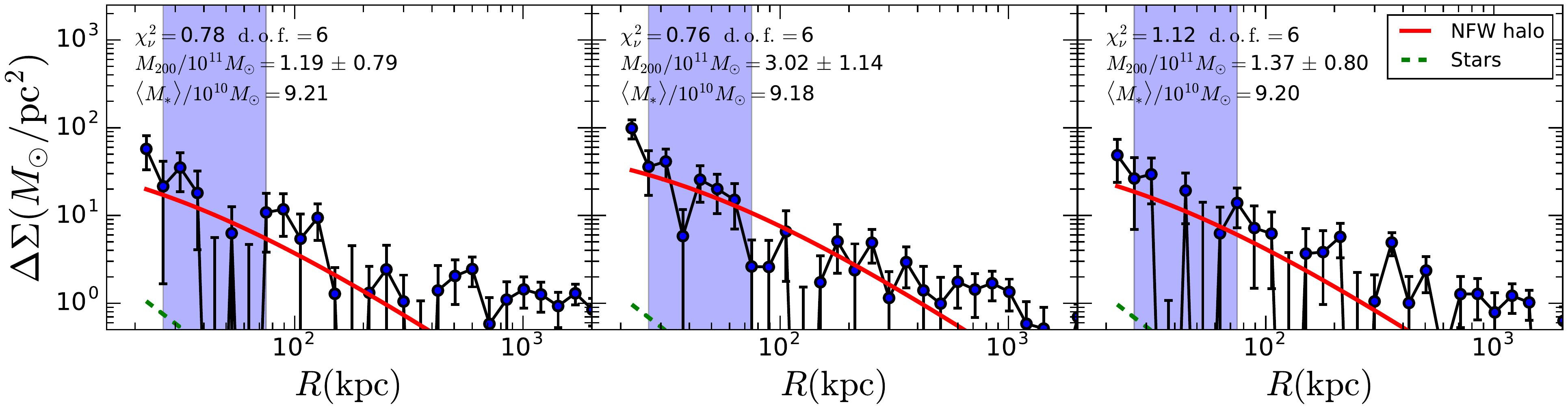}}\newline
	\subfloat{\includegraphics[width=\textwidth]{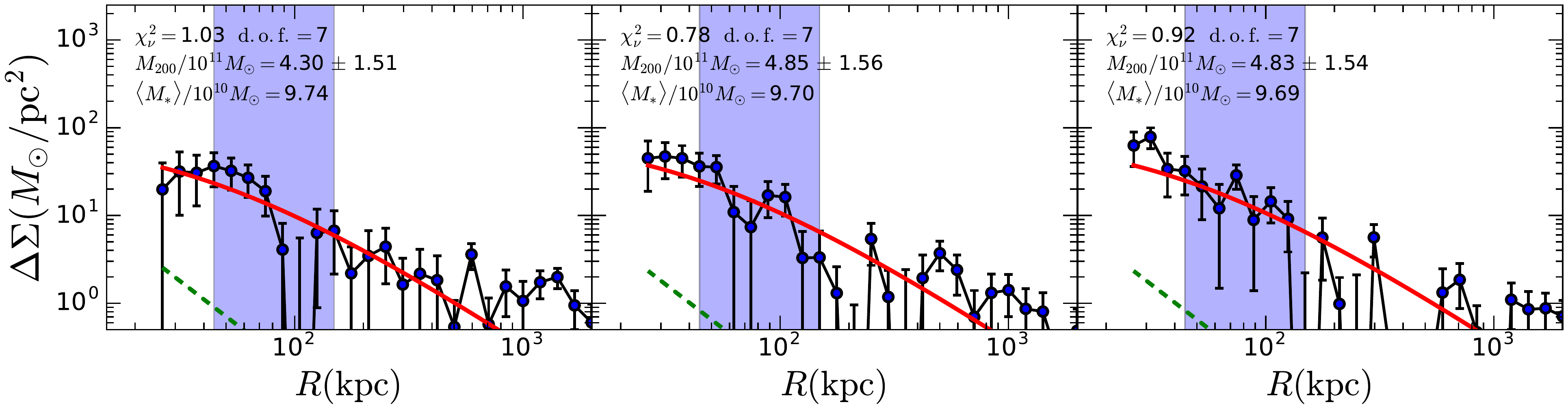}}\newline
	\subfloat{\includegraphics[width=\textwidth]{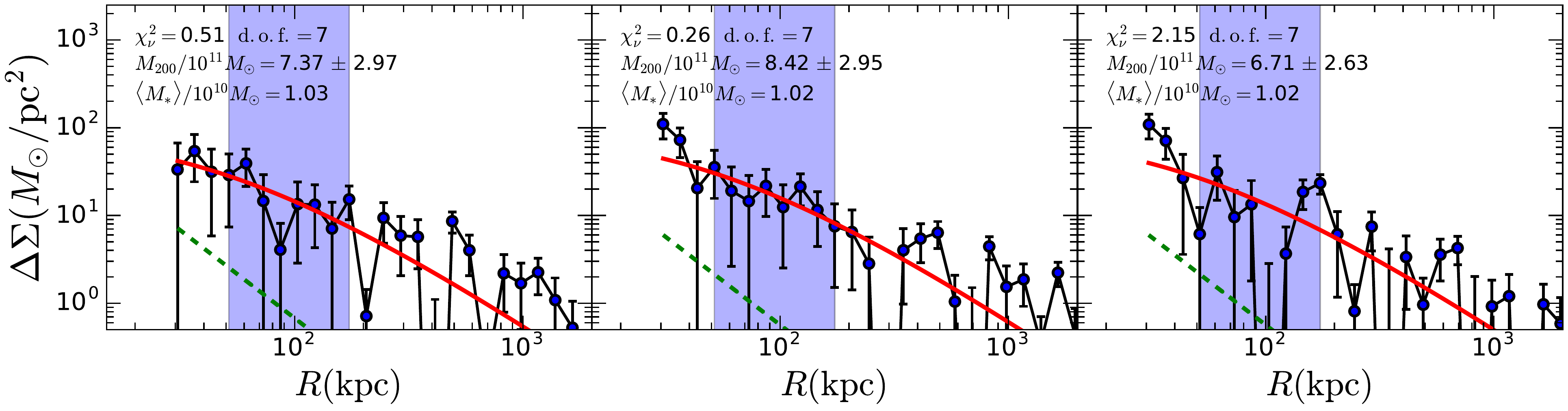}}\newline
	\subfloat{\includegraphics[width=\textwidth]{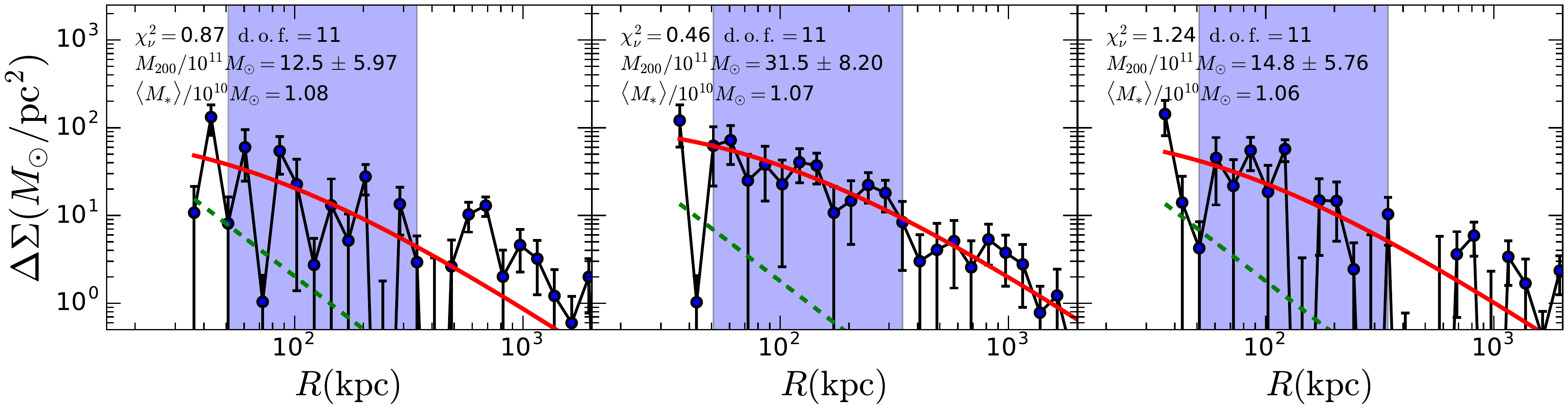}}
	\caption[Blue mid-z NFW fits]{As in Figure \ref{fig:nfw_b_z03} for blue lenses, $0.4<z<0.6$.}
	\label{fig:nfw_b_z05}
\end{figure*}

\begin{figure*}
	\subfloat{\includegraphics[width=\textwidth]{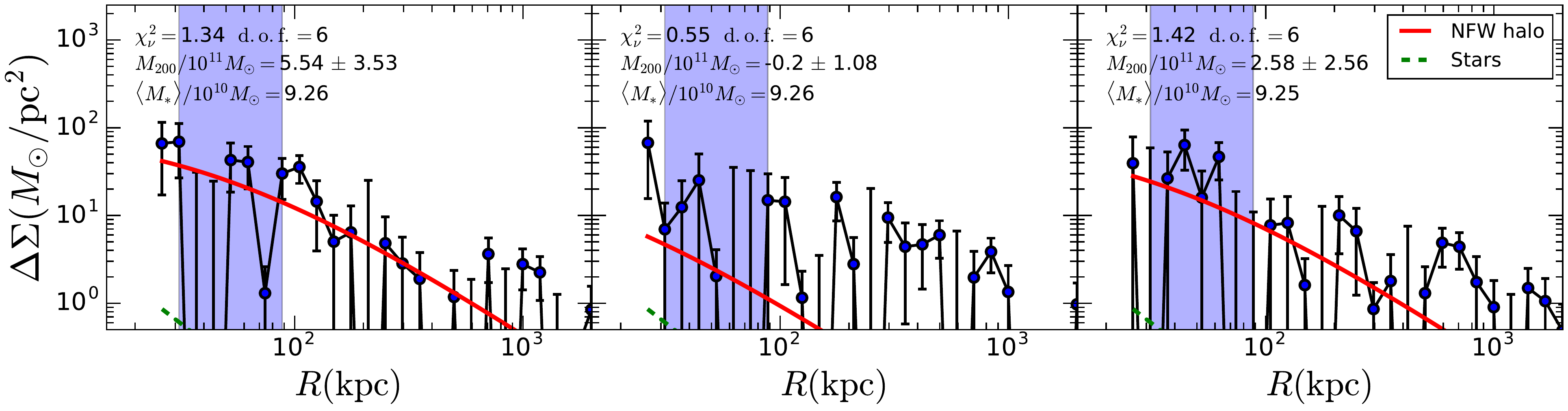}}\newline
	\subfloat{\includegraphics[width=\textwidth]{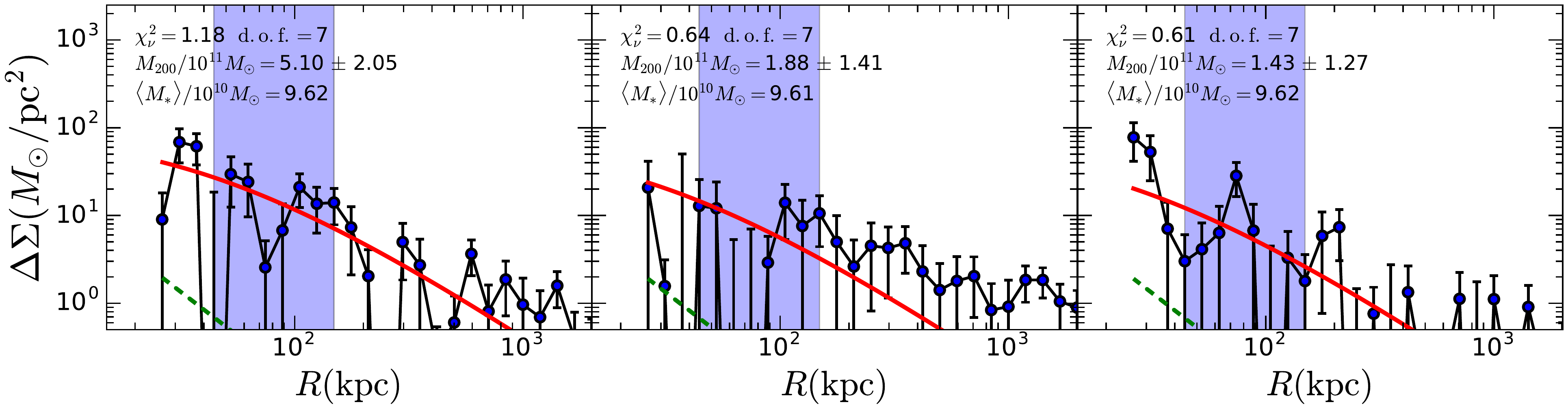}}\newline
	\subfloat{\includegraphics[width=\textwidth]{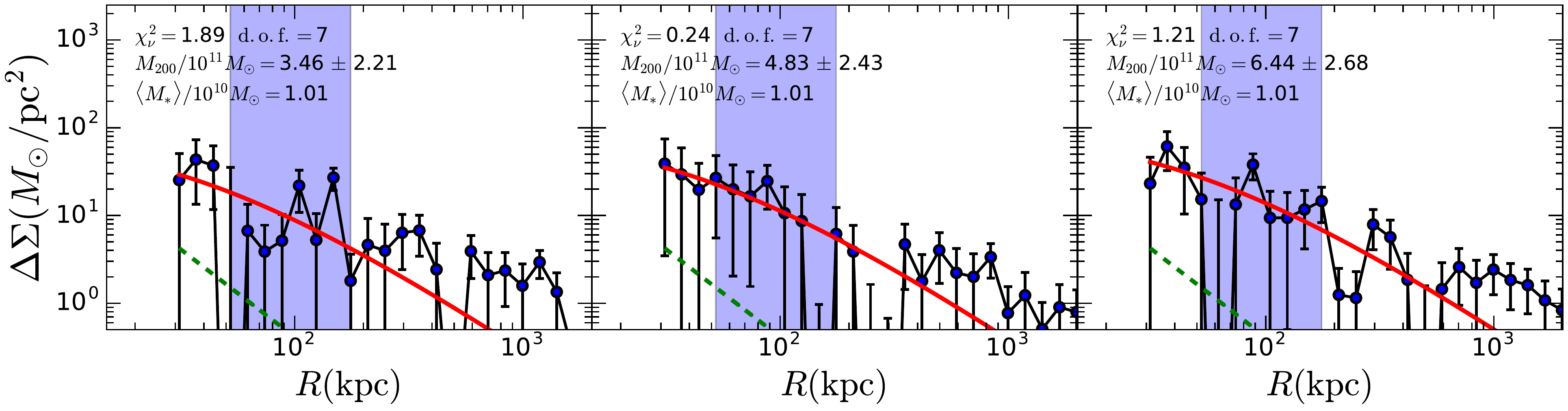}}\newline
	\subfloat{\includegraphics[width=\textwidth]{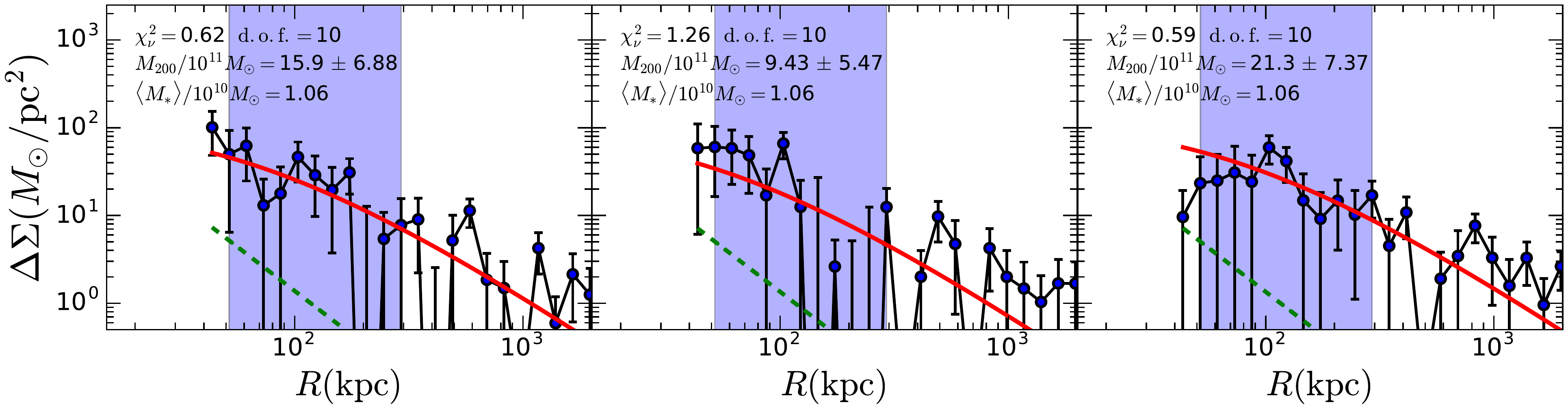}}
	\caption[Blue high-z NFW fits]{As in Figure \ref{fig:nfw_b_z03} for blue lenses, $0.6<z<0.8$.}
	\label{fig:nfw_b_z07}
\end{figure*}

\begin{figure*}
	\subfloat{\includegraphics[width=\textwidth]{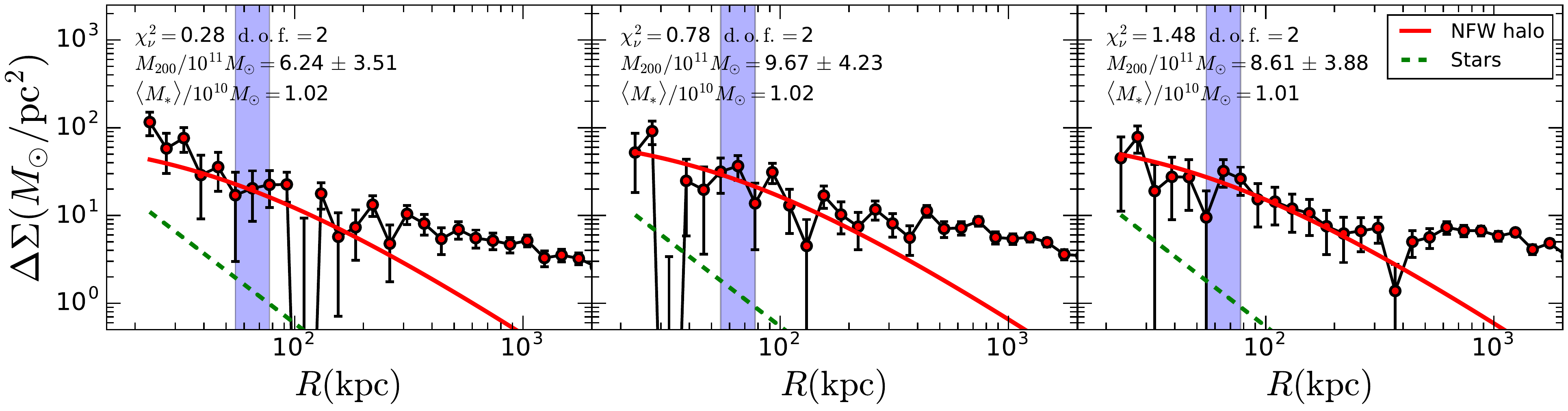}}\newline
	\subfloat{\includegraphics[width=\textwidth]{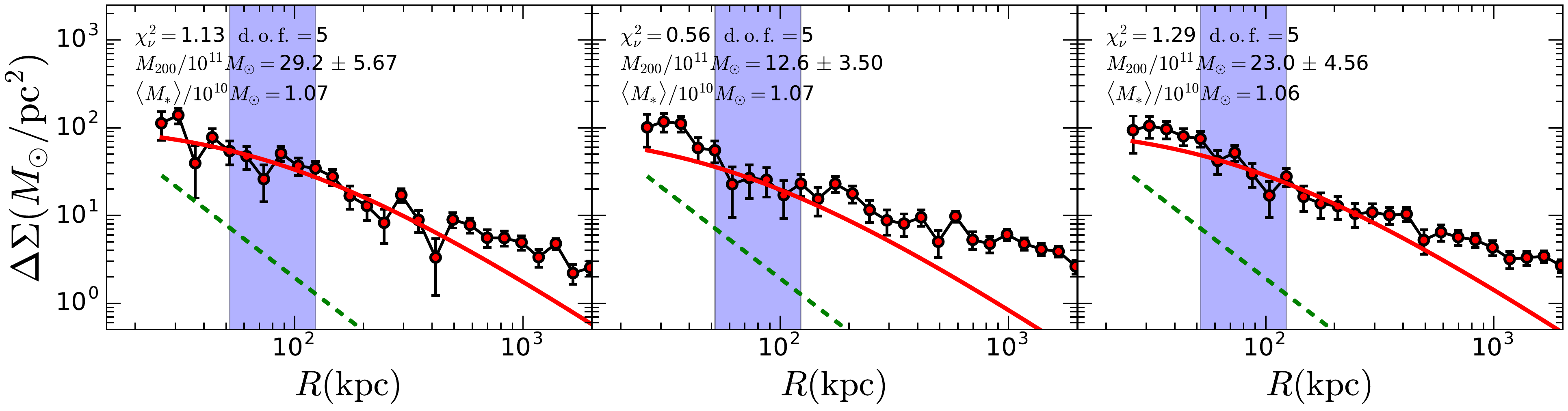}}\newline
	\subfloat{\includegraphics[width=\textwidth]{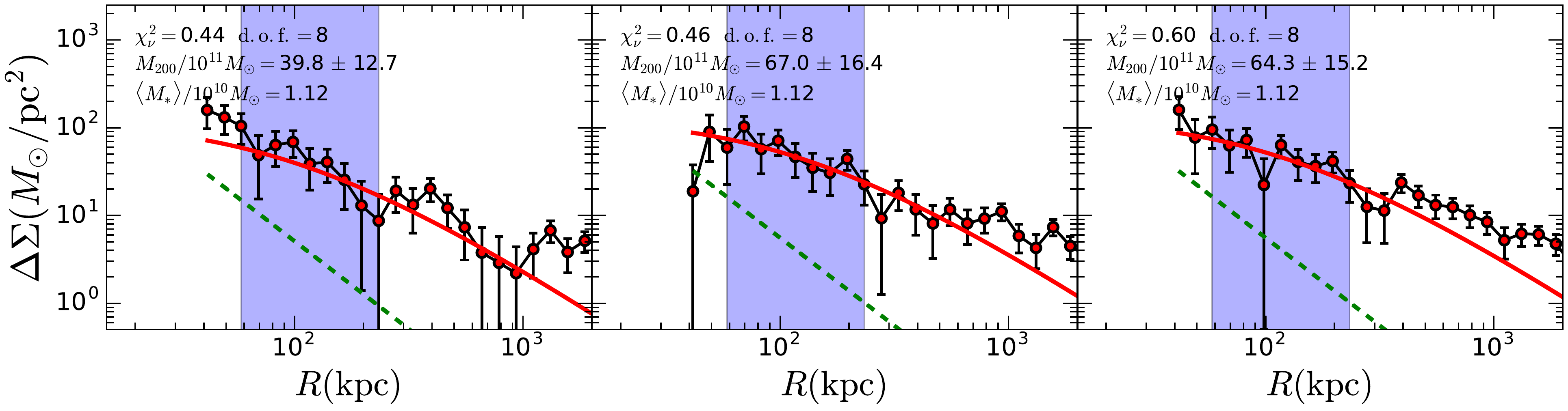}}\newline
	\subfloat{\includegraphics[width=\textwidth]{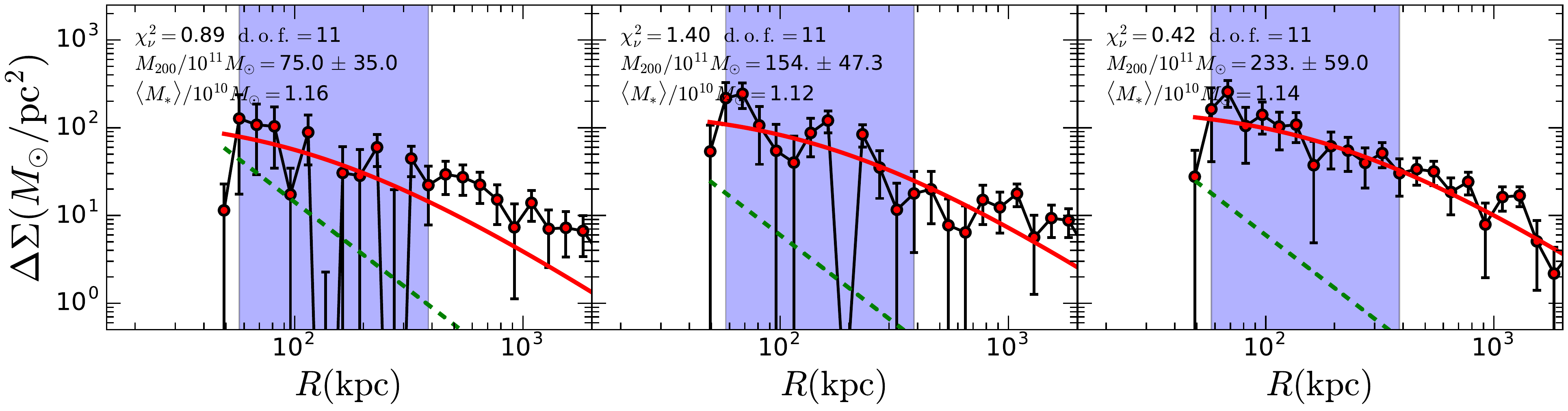}}
	\caption[Red low-z NFW fits]{As in Figure \ref{fig:nfw_b_z03} for red lenses, $0.2<z<0.4$.}
	\label{fig:nfw_r_z03}
\end{figure*}

\begin{figure*}
	\subfloat{\includegraphics[width=\textwidth]{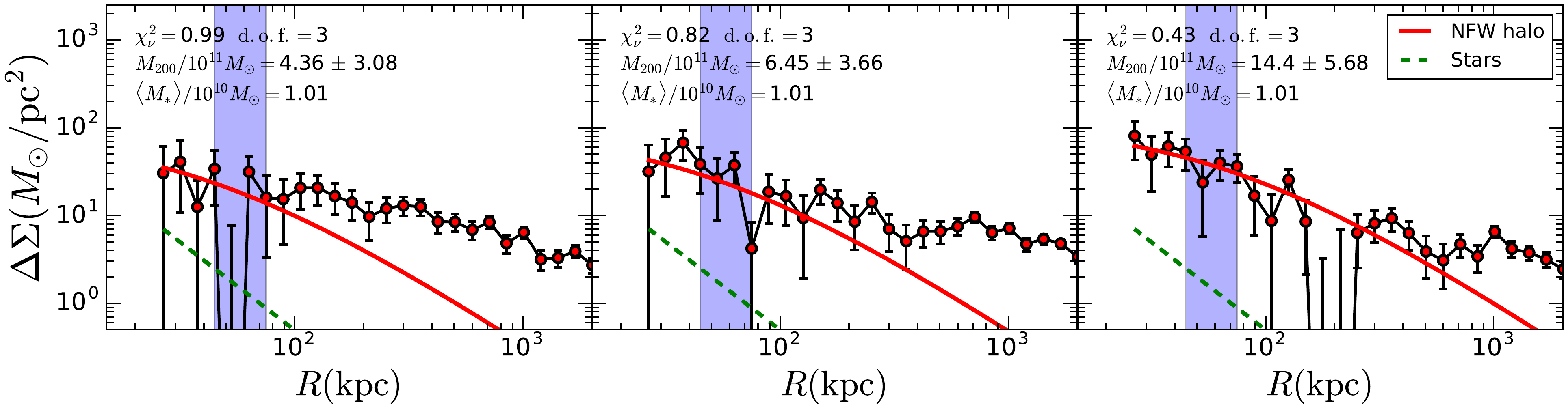}}\newline
	\subfloat{\includegraphics[width=\textwidth]{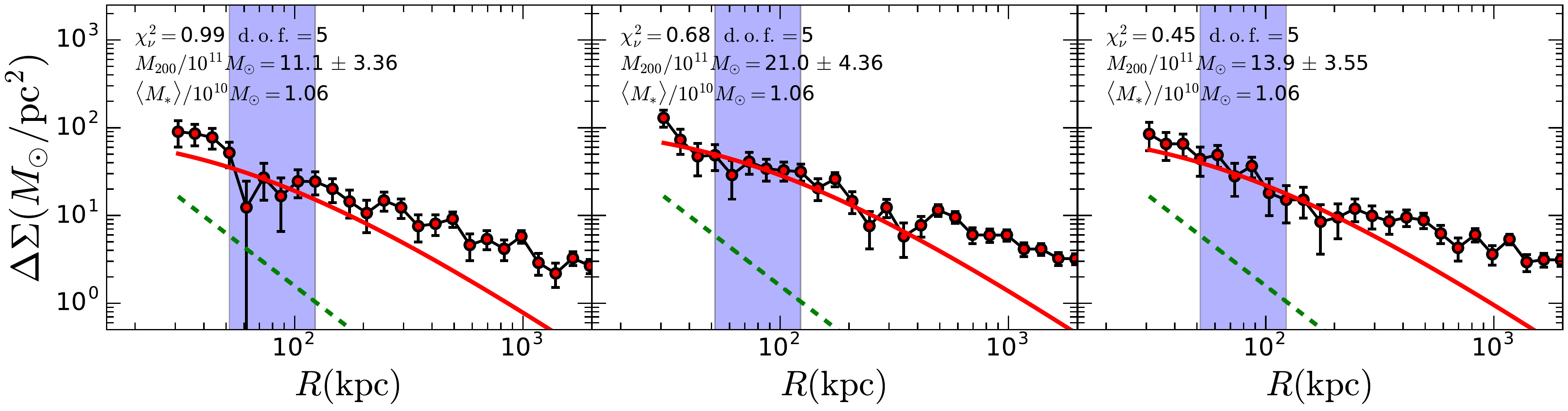}}\newline
	\subfloat{\includegraphics[width=\textwidth]{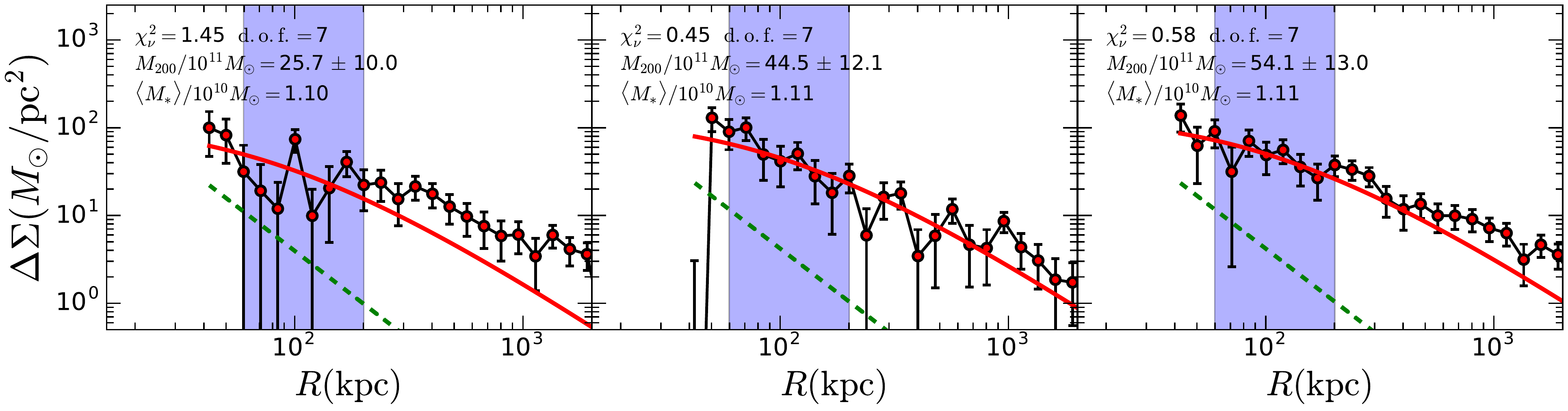}}\newline
	\subfloat{\includegraphics[width=\textwidth]{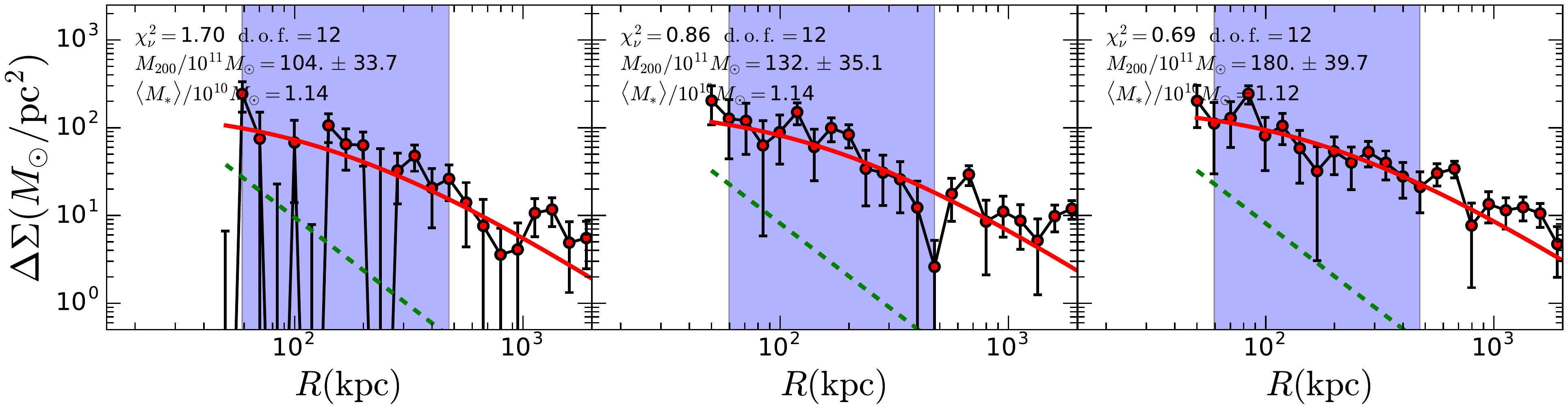}}
	\caption[Red mid-z NFW fits]{As in Figure \ref{fig:nfw_b_z03} for red lenses, $0.4<z<0.6$.}
	\label{fig:nfw_r_z05}
\end{figure*}

\begin{figure*}
	\subfloat{\includegraphics[width=\textwidth]{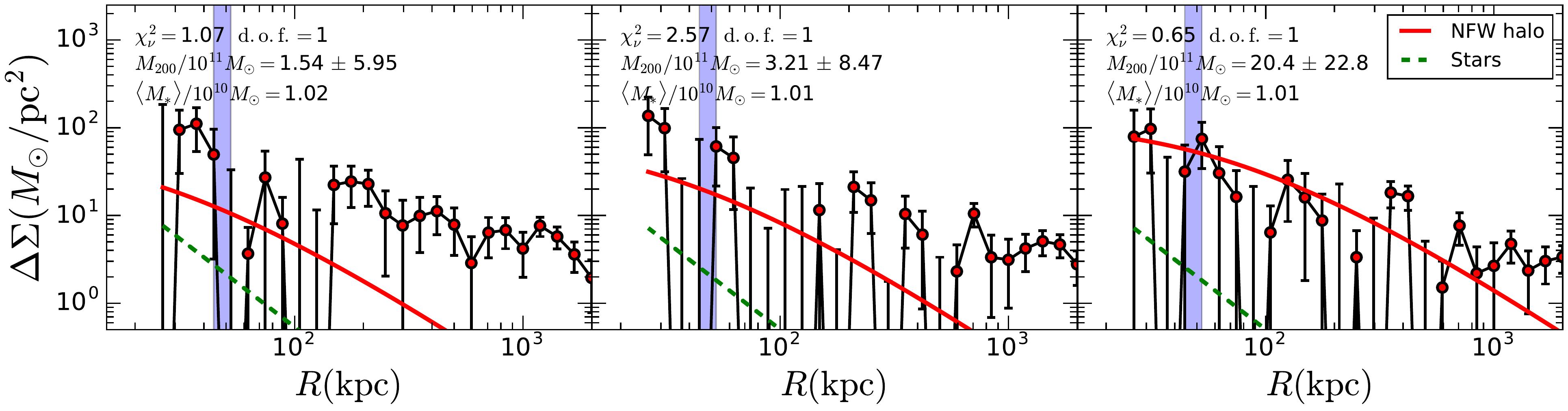}}\newline
	\subfloat{\includegraphics[width=\textwidth]{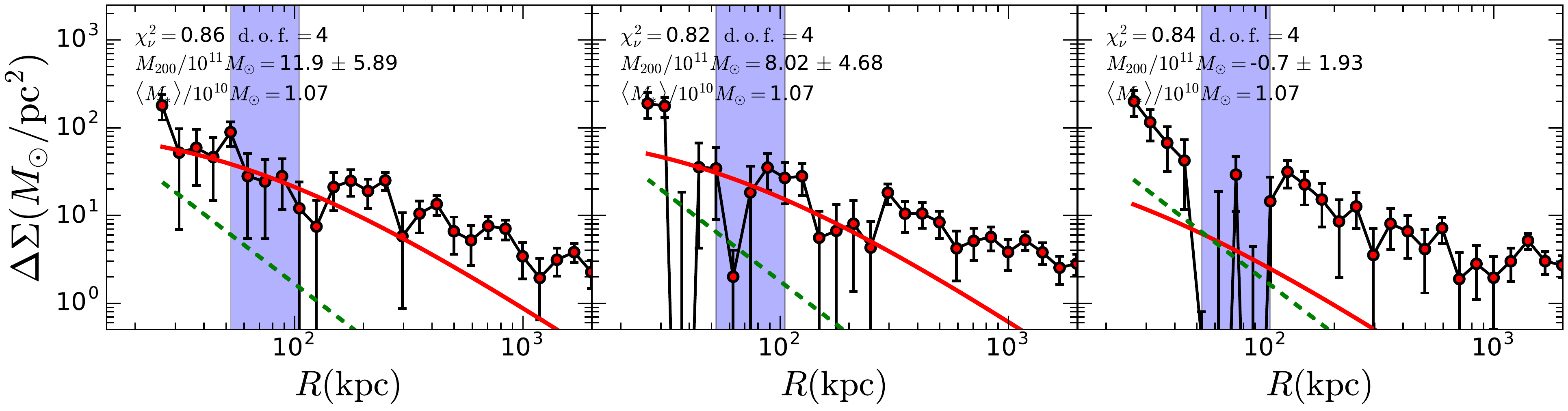}}\newline
	\subfloat{\includegraphics[width=\textwidth]{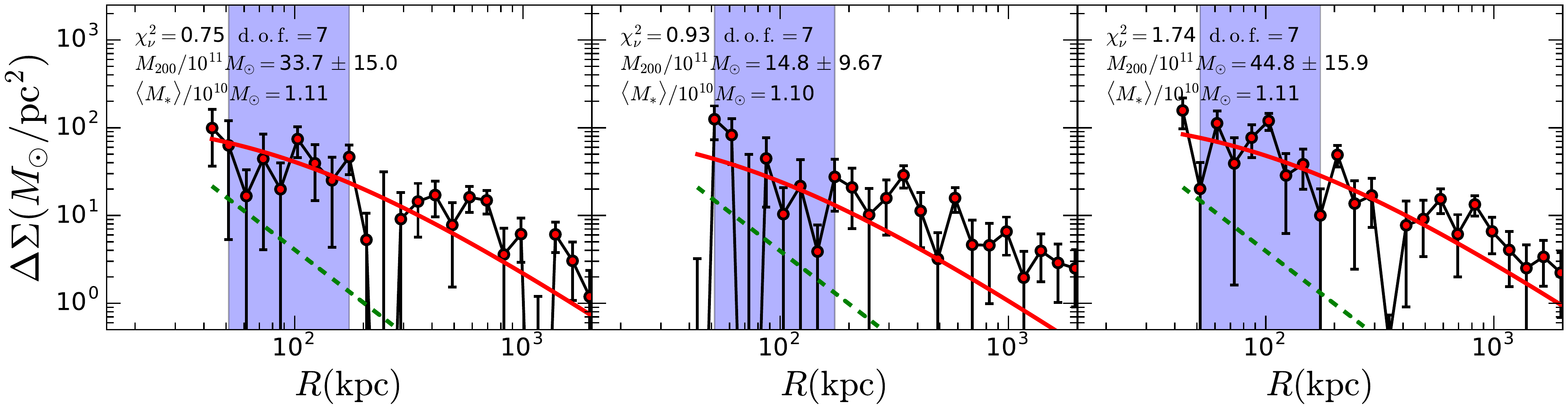}}\newline
	\subfloat{\includegraphics[width=\textwidth]{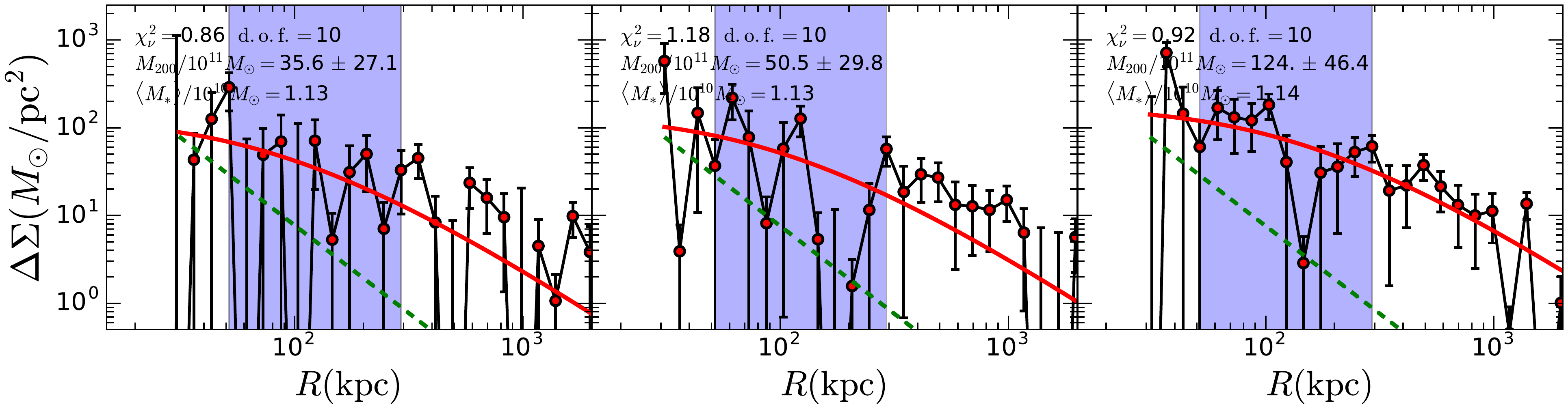}}
	\caption[Red high-z NFW fits]{As in Figure \ref{fig:nfw_b_z03} for red lenses, $0.6<z<0.8$. Due to the small fitting range the low mass bin was omitted from the final results, it is included here for completeness.}
	\label{fig:nfw_r_z07}
\end{figure*}

The full NFW fits to the $\Delta\Sigma$ profiles of each colour, $z_{\mathrm{p}}$, $M_{*}$, and $r_{\mathrm{eff}}$ bin are shown in Figures \ref{fig:nfw_b_z03}-\ref{fig:nfw_r_z07}. The NFW term is optimized such that the sum of the halo and stellar term fit the data points within the blue highlighted region. The stellar term is fixed by the $\langle M_{*} \rangle$ of the lenses in each size bin. The difference in $\langle M_{*} \rangle$ between each size bin is minor, and there is no systematic trend for larger galaxies to have higher/lower $\langle M_{*} \rangle$ than small galaxies.

\subsection{Exclusion of offset group term}
\label{sec:offset}
The presence of a more massive host halo around a satellite lens creates what is known as the offset group term, as described in Section \ref{sec:halomodel}, a convolution of the contributions of the more massive host halos surrounding lenses which are satellites. Care was taken to confine the NFW fits to the region where the lens' own halo was dominant over the group, or cluster halos in which they may be embedded, but the influence of the offset group term cannot be eliminated in this manner, only minimized. 

In this paper, we have not defined the NFW fitting region differently for satellite lenses of different $r_{\mathrm{eff}}$. For instance, if smaller galaxies were preferentially likely to be satellites compared to larger galaxies, our method would overestimate the NFW halo mass of smaller galaxies compared to the largest galaxies, by failing to account for the increased offset group term, and the fitted $\eta$ would be lower than the true $\eta$. Qualitative examination of the NFW fits shown in Figures \ref{fig:nfw_b_z03}-\ref{fig:nfw_r_z07} indicates no consistent excess $\Delta\Sigma$ in small or average-sized lenses relative to large lenses. These fits do have substantial noise such that this conclusion is not definitive, a higher signal-to-noise ratio would allow us to draw stronger conclusions in future examinations of this effect.

In order to estimate the magnitude of this effect, we can perform the following analysis. The strength of the offset group term depends on the satellite fraction, $f_{\mathrm{sat}}$. We can set an upper limit on $\eta$ if we assume that all of the largest 33\% galaxies in a given mass bin are centrals, and use $f_{\mathrm{sat}}$ from \cite{2015MNRAS.447..298H} for the average and small size bins ($f_{\mathrm{sat}}$ for average galaxies and $2f_{\mathrm{sat}}$ for smallest 33\% of the galaxies). The total integrated $\Delta\Sigma$ due to the group halo term relative to the total $\Delta\Sigma_{\mathrm{NFW}}$ over our fitting range allows us to estimate the relative contribution of the group term. Since the group term is only relevant for satellites, this ratio is multiplied by the revised satellite fraction in each size bin. This leads to lower fitted galaxy masses for the average and small size bins. Hence when $\eta$ is refitted using these modifications, $\eta$ is, on average, systematically 0.2-0.3 greater than shown by the analysis in this paper, but within the original uncertainties. The effect is largest for the bins that have the highest satellite fractions, namely red galaxies with lower stellar masses. The effect is likely to be negligible for blue galaxies which have low satellite fractions. As noted above, this is correction is an upper limit since, for example, not all large galaxies will be centrals.

\section{Simulation Data}
\label{sec:simdata}

\begin{table*}
	\caption{Fit results for EAGLE galaxies.}
	\begin{tabularx}{\textwidth}{l l l l l l l l l l} 
		\hspace{1cm}Colour & $\mathrm{N}$ & $\mathrm{f_{sat}}$ & $\mathrm{\log_{10}(\langle M_{*,all}\rangle)}$ & $\mathrm{\eta_{all}}$ & $\mathrm{\log_{10}(\langle M_{*,cent}\rangle)}$ & $\mathrm{\eta_{cent}}$ & $\mathrm{\log_{10}(\langle M_{*,sub}\rangle)}$ & $\mathrm{\eta_{sub}}$\\
		\hline
		\hspace{1cm}Blue & $\mathrm{4473}$ & $\mathrm{0.363}$ & $\mathrm{9.468}$ & $\mathrm{0.055\pm0.045}$ & $\mathrm{9.469}$ & $\mathrm{-0.172\pm0.033}$ & $\mathrm{9.466}$ & $\mathrm{0.60\pm0.10}$ \\
		\hspace{1cm}Blue & $\mathrm{2620}$ & $\mathrm{0.366}$ & $\mathrm{9.976}$ & $\mathrm{0.275\pm0.064}$ & $\mathrm{9.981}$ & $\mathrm{0.094\pm0.062}$ & $\mathrm{9.966}$ & $\mathrm{0.57\pm0.11}$ \\
		\hspace{1cm}Blue & $\mathrm{1276}$ & $\mathrm{0.290}$ & $\mathrm{10.439}$ & $\mathrm{0.51\pm0.10}$ & $\mathrm{10.447}$ & $\mathrm{0.34\pm0.10}$ & $\mathrm{10.420}$ & $\mathrm{0.68\pm0.18}$ \\
		\hspace{1cm}Blue & $\mathrm{341}$ & $\mathrm{0.173}$ & $\mathrm{10.947}$ & $\mathrm{0.78\pm0.18}$ & $\mathrm{10.959}$ & $\mathrm{0.55\pm0.17}$ & $\mathrm{10.887}$ & $\mathrm{1.22\pm0.40}$ \\
		\hspace{1cm}Red & $\mathrm{534}$ & $\mathrm{0.670}$ & $\mathrm{10.239}$ & $\mathrm{0.73\pm0.23}$ & $\mathrm{10.281}$ & $\mathrm{0.39\pm0.20}$ & $\mathrm{10.216}$ & $\mathrm{0.65\pm0.29}$ \\
		\hspace{1cm}Red & $\mathrm{478}$ & $\mathrm{0.395}$ & $\mathrm{10.771}$ & $\mathrm{0.74\pm0.11}$ & $\mathrm{10.811}$ & $\mathrm{0.400\pm0.097}$ & $\mathrm{10.700}$ & $\mathrm{0.76\pm0.27}$ \\
		\hspace{1cm}Red & $\mathrm{57}$ & $\mathrm{0.263}$ & $\mathrm{11.318}$ & $\mathrm{0.74\pm0.32}$ & $\mathrm{11.341}$ & $\mathrm{0.29\pm0.23}$ & $\mathrm{11.246}$ & $\mathrm{1.0\pm1.0}$ \\
		\hspace{1cm}Red & $\mathrm{14}$ & $\mathrm{0.000}$ & $\mathrm{11.745}$ & $\mathrm{0.06\pm0.34}$ & $\mathrm{11.745}$ & $\mathrm{0.06\pm0.34}$ & $\mathrm{--}$ & $\mathrm{--\pm--}$ \\
	\end{tabularx}
	\label{tab:eagle_results}
\end{table*}
\begin{table*}
	\caption{Fit results for Illustris galaxies.}
	\begin{tabularx}{\textwidth}{l l l l l l l l l l} 
		\hspace{1cm}Colour & $\mathrm{N}$ & $\mathrm{f_{sat}}$ & $\mathrm{\log_{10}(\langle M_{*,all}\rangle)}$ & $\mathrm{\eta_{all}}$ & $\mathrm{\log_{10}(\langle M_{*,cent}\rangle)}$ & $\mathrm{\eta_{cent}}$ & $\mathrm{\log_{10}(\langle M_{*,sub}\rangle)}$ & $\mathrm{\eta_{sub}}$\\
		\hline
		\hspace{1cm}Blue & $\mathrm{11698}$ & $\mathrm{0.261}$ & $\mathrm{9.227}$ & $\mathrm{0.504\pm 0.033}$ & $\mathrm{9.212}$ & $\mathrm{0.213\pm 0.028}$ & $\mathrm{9.269}$ & $\mathrm{0.800\pm 0.078}$ \\
		\hspace{1cm}Blue & $\mathrm{5462}$ & $\mathrm{0.267}$ & $\mathrm{9.724}$ & $\mathrm{0.450\pm 0.048}$ & $\mathrm{9.711}$ & $\mathrm{0.030\pm 0.045}$ & $\mathrm{9.758}$ & $\mathrm{0.80\pm 0.11}$ \\
		\hspace{1cm}Blue & $\mathrm{2405}$ & $\mathrm{0.257}$ & $\mathrm{10.190}$ & $\mathrm{0.342\pm 0.067}$ & $\mathrm{10.182}$ & $\mathrm{-0.124\pm 0.066}$ & $\mathrm{10.212}$ & $\mathrm{0.99\pm 0.15}$ \\
		\hspace{1cm}Blue & $\mathrm{1202}$ & $\mathrm{0.250}$ & $\mathrm{10.681}$ & $\mathrm{0.567\pm 0.099}$ & $\mathrm{10.670}$ & $\mathrm{0.17\pm 0.10}$ & $\mathrm{10.710}$ & $\mathrm{1.05\pm 0.22}$ \\
		\hspace{1cm}Red & $\mathrm{207}$ & $\mathrm{0.831}$ & $\mathrm{10.015}$ & $\mathrm{0.11\pm 0.38}$ & $\mathrm{10.028}$ & $\mathrm{-0.45\pm 0.52}$ & $\mathrm{10.012}$ & $\mathrm{0.12\pm 0.34}$ \\
		\hspace{1cm}Red & $\mathrm{225}$ & $\mathrm{0.596}$ & $\mathrm{10.653}$ & $\mathrm{0.65\pm 0.33}$ & $\mathrm{10.683}$ & $\mathrm{-0.02\pm 0.25}$ & $\mathrm{10.631}$ & $\mathrm{0.77\pm 0.51}$ \\
		\hspace{1cm}Red & $\mathrm{117}$ & $\mathrm{0.359}$ & $\mathrm{11.038}$ & $\mathrm{0.78\pm 0.27}$ & $\mathrm{11.044}$ & $\mathrm{0.29\pm 0.22}$ & $\mathrm{11.028}$ & $\mathrm{1.06\pm 0.55}$ \\
		\hspace{1cm}Red & $\mathrm{105}$ & $\mathrm{0.276}$ & $\mathrm{11.331}$ & $\mathrm{0.58\pm 0.26}$ & $\mathrm{11.340}$ & $\mathrm{0.29\pm 0.24}$ & $\mathrm{11.307}$ & $\mathrm{-0.12\pm 0.94}$ \\
		\hspace{1cm}Red & $\mathrm{38}$ & $\mathrm{0.132}$ & $\mathrm{11.776}$ & $\mathrm{0.50\pm 0.31}$ & $\mathrm{11.792}$ & $\mathrm{0.40\pm 0.37}$ & $\mathrm{--}$ & $\mathrm{--\pm--}$ \\
	\end{tabularx}
	\label{tab:illustris_results}
\end{table*}

\begin{figure*}
	\includegraphics[width=\textwidth]{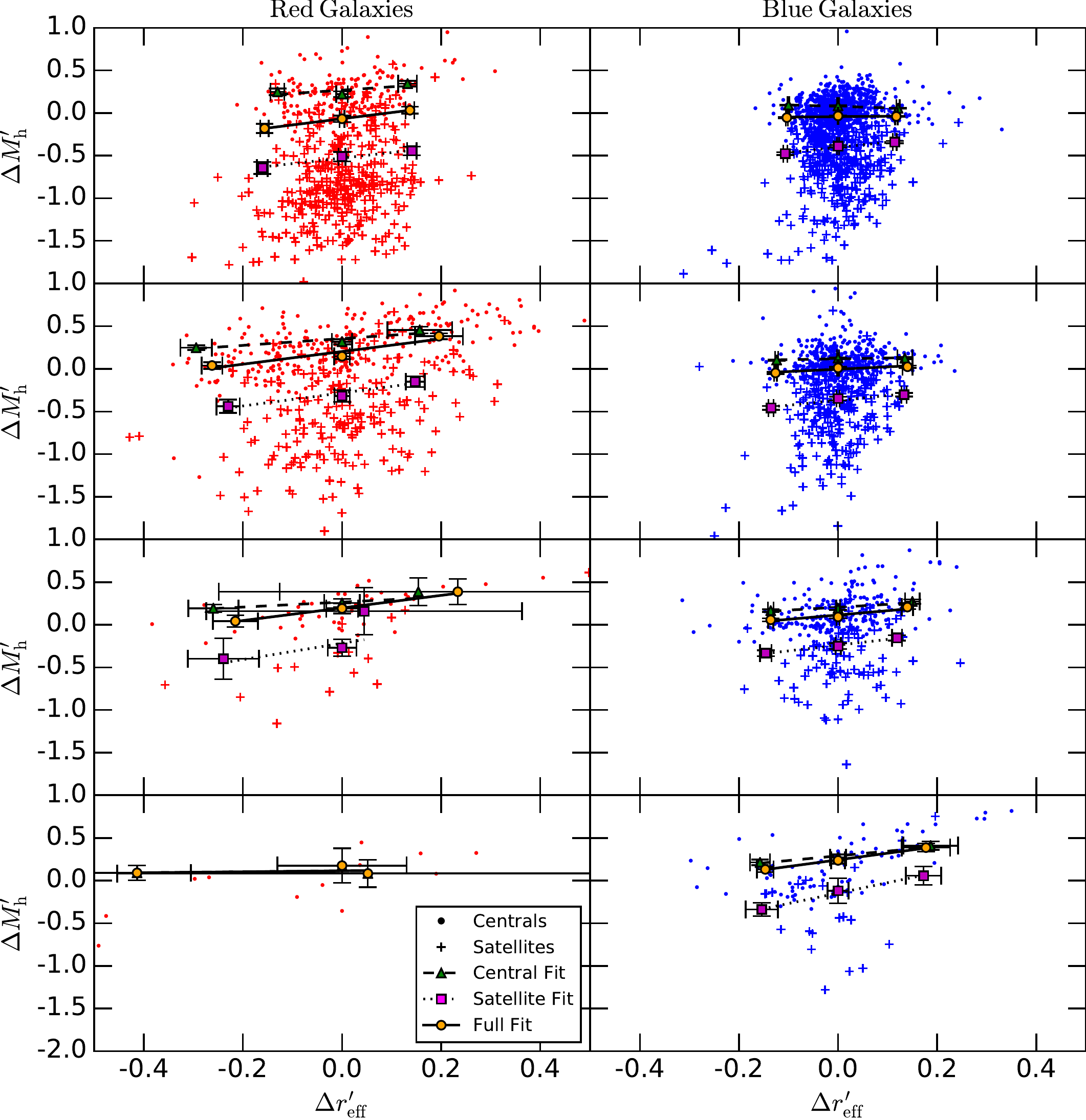}
	\caption{Halo mass-size relationships for a sample of EAGLE galaxies colour coded as red/blue galaxies, with the stellar mass bin increasing from top to bottom. Circles are central galaxies, crosses are satellites. The green, magenta, and orange points (for centrals, satellites, and the full bin, respectively) are the averages and errors on the averages that we use to fit $\eta$. The dashed, dotted, and solid lines show the best fit slopes for $\eta$, for centrals, satellites, and the full bin. The highest mass red bin (lower left) has few galaxies and is not plotted in Figure ??}
	\label{fig:eagle_ratios}
\end{figure*}
\begin{figure*}
	\includegraphics[width=\textwidth]{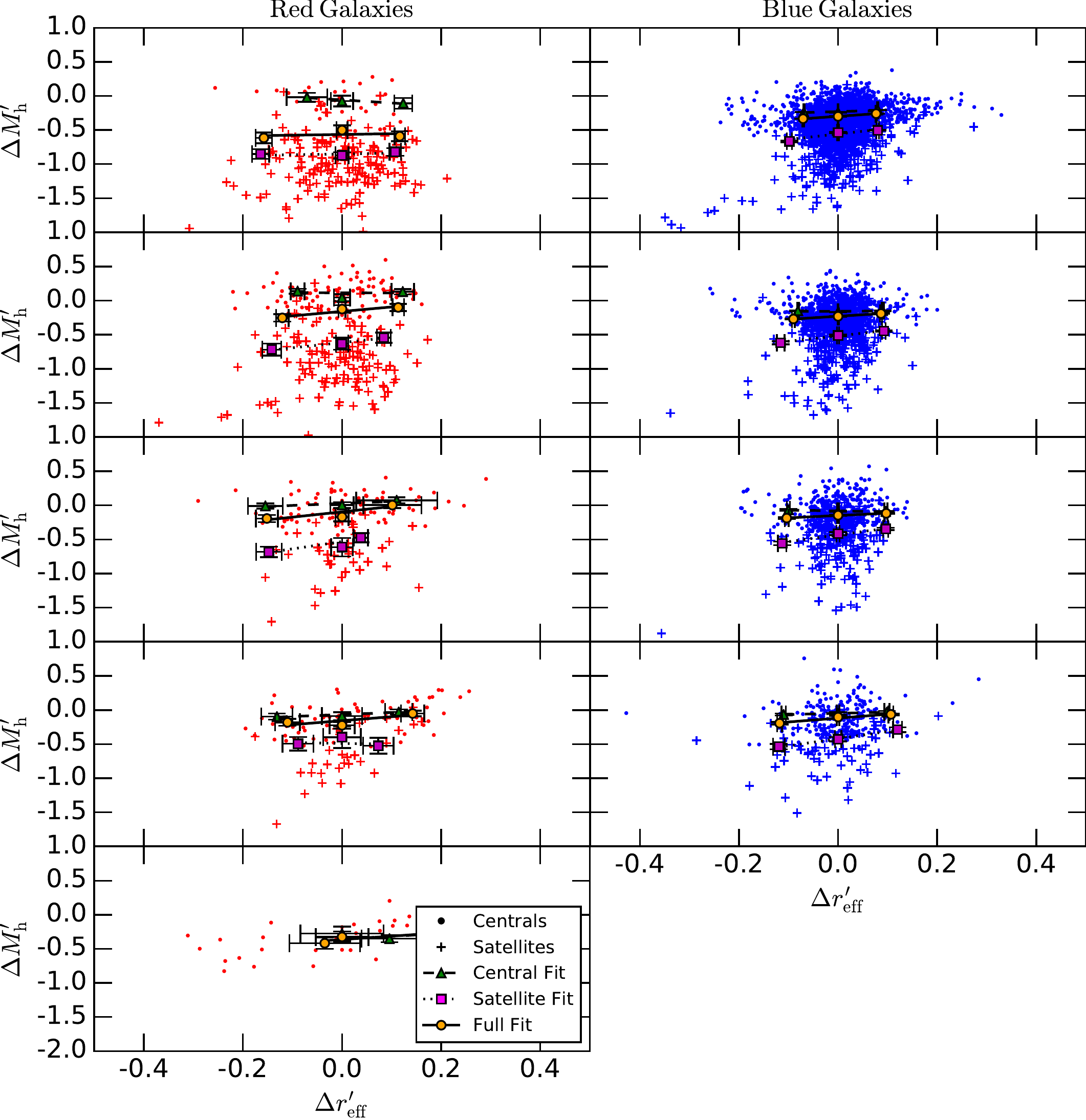}
	\caption{Halo mass-size relationships for a sample of Illustris galaxies colour coded as red/blue galaxies, with the stellar mass bin increasing from top to bottom. Circles are central galaxies, crosses are satellites. The green, magenta, and orange points (for centrals, satellites, and the full bin, respectively) are the averages and errors on the averages that we use to fit $\eta$. The dashed, dotted, and solid lines show the best fit slopes for $\eta$, for centrals, satellites, and the full bin. The highest mass red bin (lower left) has few galaxies and is not plotted in Figure ??}
	\label{fig:illustris_ratios}
\end{figure*}

In this section we include the data from the EAGLE and Illustris simulations. Plots of individual galaxies (Figures \ref{fig:eagle_ratios} \& \ref{fig:illustris_ratios}) show the differences between centrals and satellites leading to the differing $\eta$ values we fit. Centrals consist of galaxies that the simulations flag as the most massive in their subhalo groups. This means that centrals can be field galaxies with a several (or zero) satellites, brightest group galaxies, or brightest cluster galaxies. In contrast, satellites are considered to be any galaxies besides the most massive member of the main subhalo group they reside in; i.e. the second most massive galaxy in a cluster is still considered a satellite even if it has its own satellites. 

When we examine the halo mass-size plots in detail we see that both simulations are qualitatively similar. Central galaxies are more tightly clustered at higher masses and distinct from the broad swath of satellite galaxies at lower masses. For satellite galaxies, we see the same clustering of galaxies at intermediate sizes with a tail leading off at low mass and size. We also see that the three satellite galaxy bins tend toward smaller sizes, while the three central galaxy bins tend toward larger sizes. Despite overall differences in the halo-mass size relationship in both simulations, we see a consistency in the distribution of galaxies that allows us to use them as a meaningful point of comparison to our data.

\bsp	
\label{lastpage}
\end{document}